\documentclass[12pt, draftclsnofoot, journal, letter, onecolumn]{IEEEtran}
\usepackage{graphicx}
\usepackage{epsfig}
\usepackage{latexsym}
\usepackage{amsfonts}
\usepackage{here}
\usepackage{rawfonts}
\usepackage[latin1]{inputenc}
\usepackage[T1]{fontenc}
\usepackage{calc}
\usepackage{url}
\usepackage{enumerate}
\usepackage{color}
\usepackage{amssymb}
\usepackage{upref}
\usepackage{epic,eepic}
\usepackage{times}
\usepackage{dsfont}
\usepackage{comment}
\usepackage{cite}

\usepackage{graphicx}
\usepackage[font={small}]{caption}
\usepackage[font={small}]{subcaption}

\usepackage{stfloats}
\usepackage{mathtools}
\usepackage{amsmath}

\newtheorem{theorem}{\bf{Theorem}}
\newtheorem{definition}{\bf{Definition}}
\newtheorem{corollary}{\bf{Corollary}}
\newtheorem{lemma}{\bf{Lemma}}

\ifCLASSINFOpdf
\else
\fi

\begin{document}
\title{Stochastic Geometry Modeling and Analysis of Single- and Multi-Cluster Wireless Networks}

\author{Seyed~Mohammad~Azimi-Abarghouyi,~Behrooz~Makki,\\~Martin~Haenggi,~\IEEEmembership{Fellow,~IEEE},~Masoumeh Nasiri-Kenari,~\IEEEmembership{Senior~Member,~IEEE}, and~Tommy~Svensson,~\IEEEmembership{Senior~Member,~IEEE}% <-this % stops a space
\thanks{S.M. Azimi-Abarghouyi and M. Nasiri-Kenari are with the Dep. of Electrical Engineering, Sharif
University of Technology, Tehran 11365-9363, Iran. (e-mail: azimi$\_$sm@ee.sharif.edu; mnasiri@sharif.edu). B. Makki and T. Svensson are with the Dep. of Signals and Systems, Chalmers University of Technology, 412 96 Gothenburg, Sweden. (e-mail: $\{$behrooz.makki, tommy.svensson$\}$@chalmers.se). M. Haenggi is with the Dept. of Electrical Engineering, University of Notre Dame, IN 46556, USA. (e-mail: mhaenggi@nd.edu). This work has been supported in part by the Research Office of Sharif University of Technology under grant QB960605 and by the VR Research Link Project "Green Communications" and by the U.S.~National Science Foundation under grant CCF 1525904.}% <-this % stops a space
 
 }

\maketitle

\vspace{-25pt}
\begin{abstract}
This paper develops a stochastic geometry-based approach for the modeling and analysis of single- and multi-cluster wireless networks. We first define finite homogeneous Poisson point processes to model the number and locations of the transmitters in a confined region as a single-cluster wireless network. We study the coverage probability for a reference receiver for two strategies; closest-selection, where the receiver is served by the closest transmitter among all transmitters, and uniform-selection, where the serving transmitter is selected randomly with uniform distribution. Second, using Matern cluster processes, we extend our model and analysis to multi-cluster wireless networks. Here, the receivers are modeled in two types, namely, closed- and open-access. Closed-access receivers are distributed around the cluster centers of the transmitters according to a symmetric normal distribution and can be served only by the transmitters of their corresponding clusters. Open-access receivers, on the other hand, are placed independently of the transmitters and can be served by all transmitters. In all cases, the link distance distribution and the Laplace transform (LT) of the interference are derived. We also derive closed-form lower bounds on the LT of the interference for single-cluster wireless networks. The impact of different parameters on the performance is also investigated.

\end{abstract}
\begin{IEEEkeywords}
Stochastic geometry, coverage probability, clustered wireless networks, Poisson point process, Matern cluster process.
\end{IEEEkeywords}

\section{Introduction}

Single-cluster wireless networks are composed of a number of nodes distributed inside a finite region. This spatial setup is an appropriate model for, e.g., various millimeter wave communications use-case scenarios, indoor and ad hoc networks, as promising candidate technologies for the next generation of wireless networks [1]-[2]. This setup is also useful in situations where there is a range limit for backhaul links in connecting transmitters to a core network, e.g., cloud radio access networks [3]. On the other hand, the ever-growing randomness and irregularity in the locations of nodes in a wireless network has led to a growing interest in the use of stochastic geometry and Poisson point processes (PPPs) for accurate, flexible, and tractable spatial modeling and analysis [4]-[9]. 

In comparison to wireless networks on infinite regions that are mostly modeled by the infinite homogeneous PPP (HPPP) [10, Def. 2.8], the modeling and performance analysis of single-cluster wireless networks is more challenging and requires different approaches. The main challenge is that a finite point process is not statistically similar from different locations, and therefore, the system performance depends on the receiver location even after averaging over the point process.

The stochastic geometry-based modeling and analysis of single-cluster wireless networks modeled as a binomial point process (BPP) [10, Def. 2.11] has been well studied [11]-[21]. In the BPP model, a fixed and finite number of nodes are distributed independently and uniformly inside a finite region. Most prior works focus on a setup where the reference receiver is placed at the center of a circular network region [11]-[15]. Also, considering a circular region, [16] has recently developed a comprehensive framework for performance characterizations of an arbitrarily-located reference receiver inside the region under different transmitter selection strategies. Considering transmitters at a non-zero fixed altitude, disk-shaped networks of unmanned aerial vehicles are analyzed in [17]. There are also a few studies that present outage probability characterizations of a fixed link inside an arbitrarily-shaped finite region [18]-[21]. 

In spite of the usefulness of HPPP for modeling and analysis of coverage-centric and uniform deployments of nodes [4]-[10], it cannot accurately model user-centric and content-centric deployments, where the nodes may be deployed at the places with high user density [22]-[24]. In such deployments, it is important to take into account non-uniformity as well as the correlation that may exist between the locations of the transmitters and receivers. Accordingly, third generation partnership project (3GPP) has considered clustered models in [24]-[25]. Models based on Poisson cluster processes (PCPs) [10, Sec. 3.4] have recently been studied for heterogeneous [22] and device-to-device (D2D) networks [23]. In these works, the network follows a Thomas cluster process (TCP) [10, Def. 3.5]. A PCP model is also proposed and analyzed for heterogeneous networks in [26]. In [27], clustered ad hoc networks are modeled using the Matern cluster process (MCP) [10, Def. 3.6] and the TCP, and the performance of a fixed link is analyzed. Nearest-neighbor and contact distance distributions for the MCP are derived in [28].

In this paper, we develop tractable models for single- and multi-cluster wireless networks. We define a finite homogeneous Poisson point process (FHPPP) to model nodes in a finite region and then develop a framework for the analysis of single-cluster wireless networks under two different strategies. The first approach is referred to as closest-selection where a reference receiver is served by the closest transmitter in the network. In the second approach, which we refer to as uniform-selection, a uniformly randomly selected transmitter is connected to the receiver. These strategies cover a broad range of requirements of wireless networks. For instance, the closest-selection approach is suitable for cellular networks, while the uniform-selection scheme is relevant to ad hoc networks. 

To model multi-cluster wireless networks consisting of different single-cluster wireless networks, we consider an MCP of transmitters. For the receivers, we consider two types, i) closed-access receivers, which are located around the cluster centers of transmitters with a symmetric normal distribution and are allowed to be served only by the transmitters of their corresponding clusters according to the closest- or uniform-selection strategy, and ii) open-access receivers, which can be served by all transmitters according to the closest-selection strategy. 

We derive exact expressions for the coverage probability of a reference receiver in single- and multi-cluster wireless networks for the different selection strategies and types of receivers. For each selection strategy and type of receiver in single- and multi-cluster wireless networks, we characterize the Laplace
transform (LT) of the interference. Moreover, as a key step for the coverage probability analysis, the distributions of the distance from the reference receiver to its serving transmitter are derived. We also derive tight closed-form lower bounds on the LT of the interference in the case of single-cluster wireless networks, which are convenient for the coverage probability analysis.

We investigate the impact of different parameters of the system models on the performance in terms of the coverage probability and spectral efficiency. In most cases, a higher path loss exponent improves the performance. However, at relatively high distances of the reference receiver to the center of single-cluster wireless networks, a higher path loss has a degrading effect on the performance. Also, an increase in the distance of the reference receiver to the center of the network decreases the chance for coverage. Our analysis reveals that there exists an optimal distance for the location of the reference receiver from the center of the network that maximizes the
coverage probability. An optimal distance is also observed for the spectral efficiency. Our evaluation also shows that, for a broad range of parameter settings, our proposed lower bounds tightly mimic the exact results on the coverage probability.

Our work is different from the state-of-the-art literature, e.g., [10]-[23], [26]-[28], from three perspectives. First, different from the BPP, which models a fixed number of nodes in a region, we consider a point process that is suitable for finite regions with a random number of nodes and allow for arbitrary receiver locations. %Second, tackling with arbitrarily-shaped regions is cumbersome [19]. 
Second, we comprehensively study multi-cluster wireless networks using the MCP. In our analysis, we also derive the contact distribution function of the MCP in a form that is significantly simpler than the one in [28, Thm. 1]. Third, we propose open-access and closed-access receivers and different transmitter selection strategies. %None of these results have been previously presented by, e.g., [10-23].

The rest of the paper is organized as follows. Section II describes the system models. Secion III proposes the transmitter selection strategies and presents the analytical results for the coverage probability of single-cluster wireless networks, including characterizations for the serving distance distributions, and the LT of the interference and its corresponding lower bounds. Section IV presents the analytical results for the coverage probability of multi-cluster wireless networks, and derives the related serving distance distributions and the LT of the interferences. Section V presents the numerical results. Finally, Section VI concludes the paper.

%The rest of the paper is organized as follows. Section II describes the finite network model and assumptions, including the transmitter selection strategies. Section III derives the serving distance distributions for finite wireless networks. Secion IV presents the analytical results for the coverage probability of finite wireless networks, including characterizations for the LT of the interference and its corresponding lower bounds. Section V proposes our model for clustered finite wireless networks and different types of receivers, and derives the related serving distance distributions, the LT of the interferences as well as the coverage probabilities. Section VI presents the numerical results, and accordingly obtains design guidelines. Finally, Section VII concludes the paper.

\section{System Model} 	
In this section, we provide a mathematical model of the system, including the spatial distribution of the nodes for single- and multi-cluster wireless networks and the channel model.
\vspace{-30pt}
\subsection{Spatial Model for Single-Cluster Wireless Networks}
Let us define an FHPPP as follows. 
\begin{definition}
We define the FHPPP as ${\Phi}=\mathcal{P}\cap{\cal A}$, where $\mathcal{P}$ is an HPPP of intensity $\lambda$ and ${\cal A}\subset\mathbb{R}^2$.  \hspace{416pt}\IEEEQEDclosed
\end{definition}

We consider a single-cluster wireless network as shown in Fig. 1, where the locations of active transmitters are modeled as an FHPPP. The transmitters are assumed to transmit at the same power. For simplicity and in harmony with, e.g., [11]-[17], we let ${\mathbf{\cal A}}=\mathbf{b}(\mathbf{x}_\mathbf{o},D)$, where $\mathbf{b}(\mathbf{x}_\mathbf{o},D)$ represents a disk centered at $\mathbf{x}_\mathbf{o}$ with radius $D$. However, our theoretical results can be extended to arbitrary regions $\cal A$. %Moreover, the disk with minimum area which covers a region may well approximate its performance.  

Receivers can be located everywhere in $\mathbb{R}^2$. With no loss of generality, we conduct the analysis at a reference receiver located at the origin $\mathbf{o}$. We further define $d = \|\mathbf{x}_\mathbf{o}\|$. 

The proposed setup well models a wireless network confined in a finite region, such as indoor and ad hoc networks.
\begin{figure}[tb!]
\centering

\includegraphics[width =2in]{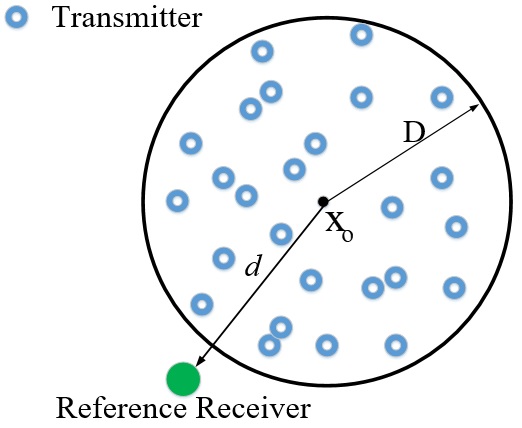}
\vspace{-5pt}
\caption{An illustration of the system model for single-cluster wireless networks.}
\vspace{-20pt}
\end{figure}

\subsection{Spatial Model for Multi-Cluster Wireless Networks}
An MCP is defined as follows [10, Def. 3.6]. 
\begin{definition}
An MCP ${\Phi}$ is a union of offspring points that are located around parent points. The parent point process is an HPPP ${\Phi}_{\rm p}$ with intensity $\lambda_{\rm p}$, and the offspring point processes (one per parent) are conditionally independent. Conditioned on $\mathbf{x}\in {\Phi}_{\rm p}$, the offsprings form an FHPPP ${\Phi}_\mathbf{x}$ with intensity $\lambda$ over the disk $\mathbf{b}(\mathbf{x},D)$.  \hspace{245pt}\IEEEQEDclosed
\end{definition}
We consider a multi-cluster wireless network as shown in Fig. 2, where the locations of active transmitters are modeled as an MCP.

We consider two types of receivers. The first type, referred to as closed-access receivers, can be served by only a single cluster of transmitters. This closed-access receiver is distributed according to a symmetric normal distribution with variance $\sigma_\text{c}^2$ around the parent point of its corresponding cluster. Therefore, assuming a closed-access receiver at $\mathbf{y}$ and its parent point at $\mathbf{x}$, $\|\mathbf{x}-\mathbf{y}\|$ is Rayleigh distributed with probability density function (PDF)
\begin{eqnarray}
{f_{\|\mathbf{x}-\mathbf{y}\|}}\left( v \right) = \frac{v}{{ {\sigma_\text{c} ^2}}}{\rm{\exp}}\left( { - \frac{{{{v}^2}}}{{2{\sigma_\text{c} ^2}}}} \right).
\end{eqnarray}
The second type, referred to as open-access receivers, considers receivers that are placed independently of the transmitters and can be served by all transmitters. 

\begin{figure}[tb!]
\centering
\includegraphics[width =3.7in]{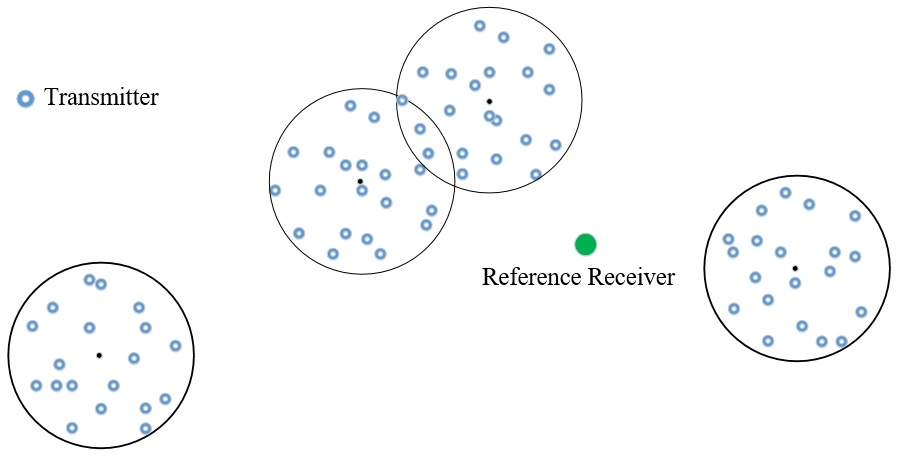}
\vspace{-5pt}
\caption{An illustration of the system model for a finite piece of multi-cluster wireless networks.}
\vspace{-20pt}
\end{figure}

The proposed setup well models various use-case scenarios as follows:

\textit{1) Clustered Small-Cell Base Stations (BSs):} The trend in cellular networks is to deploy small-cell BSs at the places with high user density, referred to as user-centric networks [22], [29], as also proposed in 3GPP [24]-[25]. In this way, according to our setup, users who are likely to be close to a cluster of small-cell BSs can be modeled as closed-access receivers, such as users at a stadium or a mall. The open-access receivers can model users who are distributed homogeneously and independently of the small-cell BS locations, such as pedestrians or cars. 

\textit{2) Cloud BSs:} A cloud BS is a distributed multiple-antenna system formed by a number of simple antenna terminals [30]. In this application, the closed-access receivers can be modeled as the users who have a license to use a certain BS, while the open-access receivers can model users with flexibility to access all BSs.

\textit{3) Clustered Access Networks:} A large building may have a number of WiFi access points as an access network to meet its users' demands. The closed-access receivers model the users who are in a building and use its access network. On the other hand, the open-access receivers are users who can handoff between access networks of different buildings.

\textit{4) Clustered D2D Networks:} A device typically has nearby devices in a finite region as a cluster in a content-centric network, which can have direct communications with each other. The closest- or uniform-selection strategy can be considered for cellular or ad hoc access to contents distributed over the devices, respectively.

\subsection{Channel Model}
We assume distance-dependent power-law path loss and
small-scale Rayleigh fading. Thus, the received power at the reference receiver from a transmitter located at $\mathbf{y}$ is ${h_\mathbf{y}}{\| \mathbf{y} \|^{ - \alpha }}$, where the (common) transmit power is set to 1 with no loss of generality and $\alpha > 2$ is the path loss exponent. The sequence $\left\{h_\mathbf{y}\right\}$ consists of i.i.d. exponential random variables with mean 1.

\section{Single-Cluster Wireless Networks}
In this section, we concentrate on single-cluster wireless networks. To allocate a transmitter to a reference receiver, we propose selection strategies in Subsection III.A. Then, distance distributions and coverage probabilities for the selection strategies are derived in Subsections III.B and III.C, respectively. However, the resulting expressions for the coverage probabilities are not very easy to use. Hence, we derive a closed-form lower bound for the coverage probability of each strategy in Subsection III.D. 

\subsection{Selection Strategies}
\textbf{1) Closest-selection.} Here, a reference receiver is served by the transmitter that provides the maximum received power averaged over the fading. In our model, this leads to the closest-selection strategy, i.e.,
\begin{eqnarray}
{{\rm{\mathbf{x}}_\text{c}}} = \arg \mathop {\min }\limits_{{\mathbf{y}} \in \left\{{\Phi} \mid n({\Phi})>0\right\}} \| {{\mathbf{y}}} \|,
\end{eqnarray}
where $n(\cdot)$ denotes the number of elements in a set. Suitable for networks with infrastructure such as downlink cellular networks, this strategy implies that a receiver is served by the transmitter whose Voronoi cell it resides in [4]-[23].

\textbf{2) Uniform-selection.} Here, the serving transmitter is selected randomly with uniform distribution among all transmitters. This leads to
\begin{eqnarray}
{{\rm{\mathbf{x}}_\text{u}}} = \text{Unif}\left\{ {\Phi}\mid n(\Phi)>0\right\},
\end{eqnarray}
where $\text{Unif}\left\{\cdot \right\}$ denotes the uniform-selection operation. Uniform-selection models random allocation of receivers to transmitters, which may be the case in networks without infrastructure, e.g., ad hoc networks and D2D networks [16], [23]. It is also suitable for applications where the content of interest for a receiver can be available at each transmitter among all transmitters with equal probability, such as caching networks.

The signal-to-interference-and-noise ratio (SINR) of a reference receiver at the origin can be expressed as
\begin{eqnarray}
\text{SINR}_q =\frac{{{h_{\mathbf{x}_q}}{{{\| \mathbf{x}_q \|}^{-\alpha}}}}}{{{\sigma ^2} + {\cal I}_q}}, \ q=\left\{\text{c},\text{u}\right\},
\end{eqnarray}
where $\mathbf{x}_q$ is the location of the serving transmitter, ${\cal I}_q = \mathop \sum \nolimits_{\mathbf{y} \in {\Phi} \backslash \left\{\mathbf{x}_q\right\}} {h_\mathbf{y}}{\| \mathbf{y} \|^{-\alpha}}$ denotes the interference, and $\sigma ^2$ is the noise power. To distinguish between the strategies, $q=\text{c}$ and $\text{u}$ represent the case of closest- and uniform-selection strategies, respectively. For notational simplicity, let us also define $R_q=\|\mathbf{x}_q\|$.
\subsection{Serving Distance Distribution}
Considering the closest- and uniform-selection strategies, in this subsection, we derive 
the distributions of the distance from the reference receiver to its serving transmitter. These distance distributions will be used later in the coverage probability analysis. 

Let us first define $\varphi_0= {\sin ^{ - 1}}\left( {\frac{D}{d}} \right)$, $\varphi_1 (r) = {\cos ^{ - 1}}\left( {\frac{{{r^2} + {d^2} - {D^2}}}{{2dr}}} \right)$, $R_1 (\theta)= d\cos \left( \theta  \right)+\\\sqrt {{D^2}-{d^2}{{\sin }^2}\left( \theta  \right) }$ and $\hat R_1 (\theta) = d\cos \left( \theta  \right)-\sqrt {D^2-{d^2}{{\sin }^2}\left( \theta  \right) }$, and present a lemma on the intersection area of two circles. 
\begin{lemma}
Consider two circles with radii $D$ and $r$ with centers separated by distance $d$. The area of their intersection is given by [31, Eq. (12.76)]
\begin{eqnarray}
{{\cal B}_d}(r)= D^2{\cos ^{ - 1}}\left( {\frac{{D^2 + {d^2} - r^2}}{{2d{D}}}} \right) + r^2 \varphi_1 (r)- \frac{1}{2}\sqrt {\left[ {{{\left( {{r} + d} \right)}^2} - D^2} \right]\left[ {D^2 - {{\left( {{r} - d} \right)}^2}} \right]}. 
\end{eqnarray}
\hspace{+459pt}\IEEEQEDclosed
\end{lemma}

Considering the closest-selection strategy, the distance from the reference receiver to its nearest transmitter $R_\text{c}$ is larger than $r$ if and only if at least one transmitter exists inside $\cal A$ and there is no transmitter located within $\mathbf{b}(\mathbf{o},r) \cap \cal A$. Letting ${\rm{\cal C}}_r$ denote the intersection, we have
\begin{eqnarray}
{\mathbb{P}}\left( {R_\text{c} > r} \right) = \frac{\mathbb{P}(n(\Phi\cap {\cal C}_r)=0 \text{ and } n(\Phi)>0)}{\mathbb{P}(n({\Phi})>0)} \nonumber\hspace{+190pt}\\\stackrel{(a)}{=}\frac{\mathbb{P}(n(\Phi\cap {\cal C}_r)=0) \mathbb{P}(n(\Phi\setminus {\cal C}_r)>0)}{\mathbb{P}(n({\Phi})>0)} \nonumber\hspace{177pt}\\\stackrel{(b)}{=} \frac{\exp\left( { - {\rm{\lambda }}\left| {\rm{\cal C}}_r \right|} \right)(1-\exp({ - {\rm{\lambda }}(\pi D^2-\left| {\rm{\cal C}}_r \right|})))}{1-\exp(- \lambda \pi D^2)} = \frac{\exp\left( { - {\rm{\lambda }}\left| {\rm{\cal C}}_r \right|} \right)-\exp({ - \lambda \pi D^2})}{1-\exp(-\lambda \pi D^2)},\hspace{-20pt}
\end{eqnarray}
where $|\cdot|$ denotes the area of a region. Also, $(a)$ follows from the fact that the numbers of points of a PPP in disjoint regions are independent, and $(b)$ is because $R_\text{c}\leq D+d$. Note when the intersection is the whole of $\cal A$, i.e., $|{\cal C}_r| = \pi D^2$, (6) is zero. 

According to Fig. 3, which shows illustrations of the intersection, there are two different cases as follows.
\begin{figure}[tb!]
\centering
\includegraphics[width =4.7in]{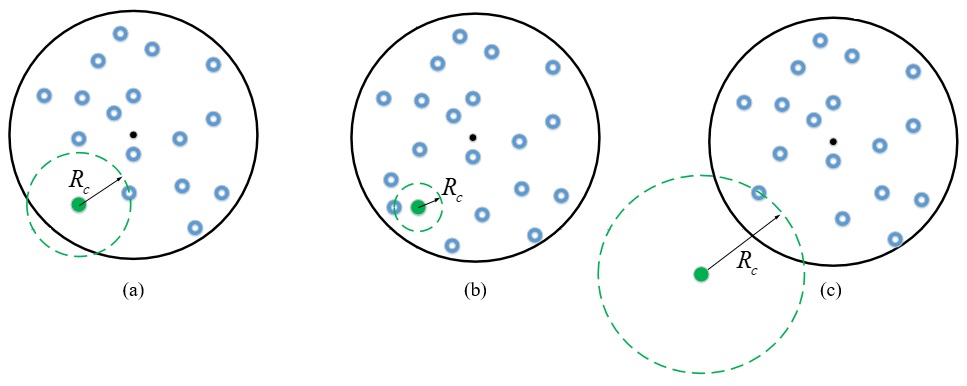}
\vspace{-10pt}
\caption{An illustration of the intersection in the case $d \leq D$ for Subplot (a): $D-d \leq R_\text{c}<D+d$ and Subplot (b): $0\leq R_\text{c}<D-d$, and in the case $d > D$ for Subplot (c): $d-D \leq R_\text{c}<d+D$. }
\vspace{-20pt}
\end{figure}
%\begin{figure}[t!]
    %\centering
    %\begin{subfigure}[t]{0.4\textwidth}
        %\centering
       %\includegraphics[width =1in]{casea.jpg}
       % \caption{$d\leq D$ and $0\leq R_\text{c}<D-d$}
    %\end{subfigure}%
    ~ %
    %\begin{subfigure}[t]{0.4\textwidth}
        %\centering
        %\includegraphics[width =1in]{caseb.jpg}
       % \caption{$d\leq D$ and $D-d \leq R_\text{c}<D+d$}
    %\end{subfigure}%
    %\begin{subfigure}[t]{0.4\textwidth}
        %\centering
        %\includegraphics[width =1in]{casec.jpg}
       % \caption{$d>D$ and $d-D \leq R_\text{c}<d+D$}
    %\end{subfigure}
   % \caption{An illustration of the intersection for the cases of $d\leq D$ and $d>D$.}
%\end{figure}

\textbf{Case 1:} If $d \le D$, then
\begin{eqnarray}
\left| {\cal C}_r \right| = \left\{ {\begin{array}{*{20}{c}}
{\pi {r^2} \hspace{+45pt} 0\leq r<D-d,}\\
{{\cal B}_d (r) \hspace{+10pt} D-d \leq r<D+d,}\\
{\pi {D^2} \hspace{+60pt} r\ge D+d,}
\end{array}} \right. 
\end{eqnarray}

where ${\cal B}_d(r)$ is given in (5).

\textbf{Case 2:} If $d > D$, then
\begin{eqnarray}
\left| {\cal C }_r\right| = \left\{ {\begin{array}{*{20}{c}}
{0 \hspace{+55pt} 0\leq r<d-D,}\\
{{\cal B}_d (r) \hspace{+10pt} d-D \leq r<D+d,}\\
{\pi {D^2} \hspace{+60pt} r\ge D+d.}
\end{array}} \right. 
\end{eqnarray}

Considering the uniform-selection strategy, on the other hand, the distance from the reference receiver to a randomly chosen transmitter, i.e., $R_\text{u}$, is less than $r$ if and only if, there is a transmitter and the transmitter is located within $\mathbf{b}(\mathbf{o},r) \cap \cal A$. Thus, as each transmitter is distributed independently and uniformly within $\cal A$, we have the following cases.

\textbf{Case 1:} If $d \le D$, then
\begin{eqnarray}
{\mathbb{P}}\left( {R_\text{u} \leq r} \right) = \left\{ {\begin{array}{*{20}{c}}
{\frac{ {r^2}}{{ D^2}} \hspace{+50pt} 0\leq r<D-d,}\\
{\frac{{\cal B}_d(r)}{{\pi D^2}} \hspace{+15pt} D-d \leq r<D+d,}\\
{1 \hspace{+78pt} r\ge D+d.}
\end{array}} \right. 
\end{eqnarray}

\textbf{Case 2:} If $d > D$, then
\begin{eqnarray}
{\mathbb{P}}\left( {R_\text{u} \leq r} \right) = \left\{ {\begin{array}{*{20}{c}}
{0 \hspace{+50pt} 0\leq r<d-D,}\\
{\frac{{\cal B}_d(r)}{{\pi D^2}} \hspace{+10pt} d-D \leq r<D+d,}\\
{1 \hspace{+75pt} r\ge D+d.}
\end{array}} \right. 
\end{eqnarray}

\subsection{Coverage Probability}
In this subsection, the distance distribution results obtained in (6)-(8) and (9)-(10) are used to derive the coverage probability of the reference receiver for the two transmitter selection strategies. As a key step in the coverage probability derivation of each strategy, we obtain the LT of the interference (Theorems 1 and 2). For notational simplicity, we define ${\cal F}(s,x)={x^2}{}_2{F_1}\left( {1,\frac{2}{\alpha };1+ \frac{2}{\alpha }; - \frac{1}{s}{{x^\alpha }}} \right)$ where ${}_2{F_1}(a,b;c;t)$ denotes the Gauss hypergeometric function [32].

\subsubsection{Closest-Selection Strategy}
\begin{theorem}
Conditioned on $R_\text{c}$, the LT of the interference under the closest-selection strategy is
\begin{eqnarray}
{\cal L}^d_{{\cal I}_\text{c}}(s|R_\text{c})=\left\{\begin{matrix}
{A}^d(s)  & \text{if} \hspace{+5pt}d\leq D  \hspace{+5pt}\text{and}  \hspace{+5pt} 0\leq R_\text{c}<D-d, \\ 
{B}^d(s)  & \hspace{+25pt}\text{if} \hspace{+5pt}d\leq D  \hspace{+5pt}\text{and}  \hspace{+5pt} D-d \leq R_\text{c}<D+d\hspace{+5pt}\\\hspace{+10pt}\text{or} \hspace{+5pt}d > D  \hspace{+5pt}\text{and}  \hspace{+5pt} \sqrt{d^2-D^2} \leq R_\text{c}<d+D,\hspace{-300pt}\\ 
{C}^d(s) & \text{if} \hspace{+5pt}d> D  \hspace{+5pt}\text{and}  \hspace{+5pt} d-D\leq R_\text{c}<\sqrt{d^2-D^2}, \hspace{-40pt}
\end{matrix}\right.
\end{eqnarray}
where ${A}^d(s)$, ${B}^d(s)$ and ${C}^d(s)$ are defined as
\begin{eqnarray}
{A}^d(s) = \exp\biggl(\pi \lambda{\cal F}(s,R_\text{c})-\lambda \int_{ 0}^{\pi}{\cal F}(s,R_1(\theta))\mathrm{d}\theta \biggr),\hspace{70pt}
\end{eqnarray}
\begin{eqnarray}
{B}^d(s) = \exp\biggl(\varphi_1(R_\text{c}) \lambda {\cal F}(s,R_\text{c}) -\lambda \int_{0}^{ \varphi_1(R_\text{c})}{\cal F}(s,R_1(\theta))\mathrm{d}\theta \biggr),\hspace{25pt}
\end{eqnarray}
\begin{eqnarray}
{C}^d(s) = B^d(s)\exp\biggl(-\lambda \int_{\varphi_1(R_\text{c})}^{\varphi_0}\Bigl\{{\cal F}(s,R_1(\theta))-{\cal F}(s,\hat R_1(\theta))\Bigr\}\mathrm{d}\theta  \biggr),
\end{eqnarray}
and $\varphi_0$, $\varphi_1$, $R_1$ and $\hat R_1$ are defined in Subsection III.B.
\end{theorem}
\begin{IEEEproof}
See Appendix A.
\end{IEEEproof}
Using the conditional LT of the interference derived in Theorem 1, we can express the coverage probability of the reference receiver for the closest-selection strategy as 
\begin{eqnarray}
{P}_{{\rm C}}^\text{c} (\beta)=\mathbb{P}(n({\Phi})>0)\mathbb{P}(\text{SINR}_\text{c} > \beta \mid n({\Phi})>0),
\end{eqnarray}
where $\beta$ is the minimum required SINR for a coverage. Note that the coverage probability is zero when there is no transmitter, and the $\text{SINR}_\text{c}$ is only meaningfully defined when there is a transmitter. Then, from (4) and averaging over the serving distance $R_\text{c}$, we have
\begin{eqnarray}
{P}_{\rm C}^\text{c} (\beta)=(1-\exp(-\lambda \pi D^2))\int_{0}^{\infty} {\mathbb{P}}\left( {\frac{{{h_{\mathbf{x}_\text{c}}}{r^{ - \alpha }}}}{{{\sigma ^2} + {\cal I}_\text{c}}} > \beta  } \right){f_{R_\text{c}}^d}\left( r \right)\mathrm{d}r.
\end{eqnarray}
where ${f_{R_\text{c}}^d}$ is the PDF of $R_\text{c}$ obtained from (6). The conditional coverage probability given a link distance $r$ can be expressed as
\begin{eqnarray}
{\mathbb{P}}\left( {\frac{{{h_{\mathbf{x}_\text{c}}}{r^{ - \alpha }}}}{{{\sigma ^2} + {\cal I}_\text{c}}} > \beta } \right) = \mathbb{P}\left( {h_{\mathbf{x}_\text{c}} > {{\beta {r^\alpha }}}\left( {{\sigma ^2} + {\cal I}_\text{c}} \right)}  \right) \stackrel{(a)}{=} \mathbb{E}\Bigl\{\exp\left(-{{{\beta {r^\alpha }}}\left( {{\sigma ^2} + {\cal I}_\text{c}} \right)}\right)\Bigr\}\nonumber\\ = \exp\left(-{{{\beta {\sigma ^2}}}} {r^\alpha }\right) {\cal L}^d_{{\cal I}_\text{c}}\left({{\beta {r^\alpha }}}|r\right),\hspace{+.6pt}
\end{eqnarray}
where $(a)$ follows from $h_{\mathbf{x}_\text{c}} \sim \exp(1)$. Finally, according to the cases considered in Subsection III.B and with $\mathcal{L}_{\cal I_\text{c}}^d$ given in (11), the coverage probability is obtained as 
\begin{eqnarray}
{P}_{\rm C}^\text{c} (\beta) =\left\{\begin{matrix}
\hspace{-30pt}\int_{0}^{D-d} 2\pi \lambda r \exp(-\lambda \pi r^2)\exp\left(-{{{\beta {\sigma ^2}}}} {r^\alpha }\right) {\cal L}_{{\cal I}_\text{c}}^d\left({{\beta {r^\alpha }}}|r\right)\mathrm{d}r \\+ \int_{D-d}^{D+d} \lambda \frac{{\partial {\cal B}_d(r) }}{{\partial r}}{\exp({ - \lambda {\cal B}_d(r) })}\exp\left(-{{{\beta {\sigma ^2}}}} {r^\alpha }\right) {\cal L}_{{\cal I}_\text{c}}^d\left({{\beta {r^\alpha }}}|r\right)\mathrm{d}r \hspace{-4pt} & \text{if} \hspace{+5pt}d \leq D, \\ 
\hspace{-10pt} \int_{d-D}^{D+d} \lambda \frac{{\partial {\cal B}_d(r) }}{{\partial r}}{\exp({ - \lambda {\cal B}_d(r) })}\exp\left(-{{{\beta {\sigma ^2}}}} {r^\alpha }\right) {\cal L}_{{\cal I}_\text{c}}^d\left({{\beta {r^\alpha }}}|r\right)\mathrm{d}r\hspace{5pt}  & \text{if} \hspace{+5pt}d > D.
\end{matrix}\right.
\end{eqnarray}
In the special case of infinite wireless networks, i.e., $D \to \infty$, the coverage probability (18) simplifies to the result in [8, Thm. 2].

\subsubsection{Uniform-Selection Strategy}
\begin{theorem}
The LT of the interference under the uniform-selection strategy is
\begin{eqnarray}
{\cal L}_{{\cal I}_\text{u}}^d (s)= \left\{\begin{matrix}
E^d(s) & \text{if} \hspace{+5pt}d \leq D, \\ 
F^d(s) & \text{if} \hspace{+5pt}d > D,
\end{matrix}\right.
\end{eqnarray}
where $E^d(s)$ and $F^d(s)$ are defined as
\begin{eqnarray}
E^d(s)= \frac{\pi D^2\exp(-\lambda \pi D^2)}{1-\exp(-\lambda \pi D^2)}\frac{\exp\Bigl(\int_{0}^{D-d}\frac{2\pi \lambda x}{{1 + s{{x}^{ - \alpha }}}}\mathrm{d}x+\int_{D-d}^{D+d}\frac{\lambda}{{1 + s{{x}^{ - \alpha }}}} \frac{{\partial {\cal B}_d(x) }}{{\partial x}}\mathrm{d}x\Bigr)-1}{\int_{0}^{D-d}\frac{2\pi x}{{1 + s{{x}^{ - \alpha }}}}\mathrm{d}x+\int_{D-d}^{D+d}\frac{1}{{1 + s{{x}^{ - \alpha }}}} \frac{{\partial {\cal B}_d(x) }}{{\partial x}}\mathrm{d}x},
\end{eqnarray}
\begin{eqnarray}
F^d(s)= \frac{\pi D^2\exp(-\lambda \pi D^2)}{1-\exp(-\lambda \pi D^2)}\frac{\exp\Bigl( \int_{d-D}^{D+d}\frac{\lambda \frac{{\partial {\cal B}_d(x) }}{{\partial x}}}{{1 + s{{x}^{ - \alpha }}}} \mathrm{d}x\Bigr)-1}{\int_{d-D}^{D+d}\frac{1}{{1 + s{{x}^{ - \alpha }}}} \frac{{\partial {\cal B}_d(x) }}{{\partial x}}\mathrm{d}x}.\hspace{119pt}
\end{eqnarray}
\end{theorem}
\begin{IEEEproof}
See Appendix B.
\end{IEEEproof}
Here the LT of the interference is independent of the serving distance $R_\text{u}$. Using Theorem 2 and following the same approach as in (16)-(18), the coverage probability for the uniform-selection strategy can be expressed as 
\begin{eqnarray}
{P}_\text{\rm C}^\text{u} (\beta) = \left\{\begin{matrix}
\hspace{-2pt}(1-e^{-\lambda \pi D^2})\Bigl\{\int_{0}^{D-d} \frac{{2r}}{D^2}\exp\left(-{{{\beta {\sigma ^2}}}} {r^\alpha }\right) {\cal L}_{{\cal I}_\text{u}}^d\left({{\beta {r^\alpha }}}\right)\mathrm{d}r \hspace{65pt}\\+ \int_{D-d}^{D+d} \frac{{1}}{\pi D^2} \frac{{\partial {\cal B}_d(r) }}{{\partial r}}\exp\left(-{{{\beta {\sigma ^2}}}} {r^\alpha }\right) {\cal L}_{{\cal I}_\text{u}}^d\left({{\beta {r^\alpha }}}\right)\mathrm{d}r\Bigr\}  \hspace{72pt}& \hspace{-17pt}\text{if} \hspace{+5pt}d \leq D, \\ 
\hspace{-7pt}(1-e^{-\lambda \pi D^2})\int_{d-D}^{D+d} \frac{{1}}{\pi D^2} \frac{{\partial {\cal B}_d(r) }}{{\partial r}}\exp\left(-{{{\beta {\sigma ^2}}}} {r^\alpha }\right) {\cal L}_{{\cal I}_\text{u}}^d\left({{\beta {r^\alpha }}}\right)\mathrm{d}r\hspace{32pt}  &\hspace{-17pt} \text{if} \hspace{+5pt}d > D.
\end{matrix}\right. \hspace{+0pt}
\end{eqnarray}

\subsection{Lower Bounds on Coverage Probability}
Since the results derived for the LT of the interference in Theorems 1 and 2 require intensive numerical computations, we derive tight lower bounds on the LT of the interference that lead to more tractable and useful analytical results. Then, using the bounds in (18) and (22), lower bounds on coverage probability can be also provided. The tightness of the bounds will be verified with numerical results (Fig. 7). 

%\textit{For the upper bound}, we approximate the integrals by partitioning the integration interval into M sub-intervals. For each of these sub-intervals,
%we use the lower limit in the integral. In fact, $\int\limits_{\varphi_1} ^{\varphi_2}  {f(\theta )d\theta }$ is approximated by $\sum\limits_{m = 0}^{M-1} {f(\varphi_1 +m\frac{\varphi_2 - \varphi_1 }{M} )\frac{\varphi_2 - \varphi_1 }{M}}$. 

To obtain the lower bounds, we outer bound the region $\cal A$ by a region that permits closed-form bounds on the LT of the interference. Note that using a larger region leads to an upper bound on the interference and, in turn, a lower bound on its LT. 

The outer region for the cases with $d\leq D$ and $d>D$ is shown in Fig. 4. Placing the center of the sectors at the reference receiver in case $d\leq D$, two covering half-circles with radii $d+D$ and $\sqrt{ D^2-d^2 }$ are considered. Also, in the case $d>D$, we consider the sector with radii $d+D$ and $d-D$ and the front angle $2 \varphi_0$ entangled between the two tangent lines. However, for the closest-selection strategy in the case $d>D$, we can achieve a tighter bound by the following regions for ${\cal A}\backslash \mathbf{b}(\mathbf{o},R_\text{c})$, which is the region including interfering transmitters. In the case $R_\text{c}>\sqrt{d^2-D^2}$, the sector with the front angle equal to twice the intersection angle, i.e., $2\varphi_1 (R_\text{c}) $, and radii $R_\text{c}$ and $D+d$ is considered. In the case $R_\text{c}<\sqrt{d^2-D^2}$, we consider two sectors with the front angle $2 \varphi_0$ and radii $R_\text{c}$ and $R_1(\varphi_1 (R_\text{c}))$ and with the front angle $2\varphi_1(R_\text{c})$ and radii $R_1(\varphi_1(R_\text{c}))$ and $D+d$.
%\begin{figure}[tb!]
%\centering
%\includegraphics[width =4.4in]{approximate1.jpg}
%\vspace{-5pt}
%\caption{Upper bounded regions in the cases Subplot (a): $d\leq D$ and Subplot (b): $d>D$.}
%\vspace{-15pt}
%\end{figure}
\begin{figure}[t!]
    \centering
    \begin{subfigure}[t]{0.5\textwidth}
        \centering
       \includegraphics[width =1.9in]{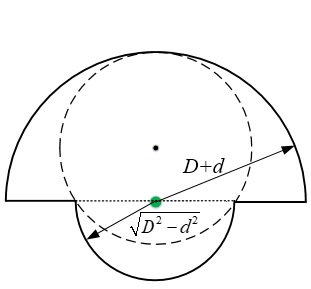}
        \caption{$d\leq D$}
    \end{subfigure}%
    ~ 
    \begin{subfigure}[t]{0.5\textwidth}
        \centering
        \includegraphics[width =1.7in]{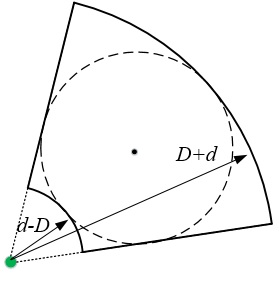}
        \caption{$d>D$}
    \end{subfigure}
     \vspace{-10pt}
    \caption{Outer bounds of $\cal A$ in the two cases $d\leq D$ and $d>D$.}
    \vspace{-25pt}
\end{figure}

In the following corollaries, we present the lower bounds on the LT of the interference for both strategies. 
%Since we have provided a general approach in details in Appendix I and II, long detailed proofs are ommited here. 
\begin{corollary}
With closest-selection, the LT of the interference given the serving distance $R_\text{c}$ is lower bounded by
\begin{eqnarray}
{\cal L}_{{\cal I}_\text{cb}}^d(s|R_\text{c})=\left\{\begin{matrix}
\hat A^d(s) &\hspace{-25pt} \text{if} \hspace{+5pt}{d\leq D}  \hspace{+5pt}\text{and}  \hspace{+5pt} {0< R_\text{c}\leq \sqrt{D^2-d^2}},\\ 
\hat B^d(s) & \text{if} \hspace{+5pt}{d\leq D}  \hspace{+5pt}\text{and}  \hspace{+5pt} {\sqrt{D^2-d^2} \leq R_\text{c}<D+d}, \\
\hat C^d(s) & \text{if} \hspace{+5pt}{d > D}  \hspace{+5pt}\text{and}  \hspace{+5pt} {d-D \leq R_\text{c}<\sqrt{d^2-D^2}},\\
\hat D^d(s) & \text{if} \hspace{+5pt}{d > D}  \hspace{+5pt}\text{and}  \hspace{+5pt} {\sqrt{d^2-D^2} \leq R_\text{c}<D+d}.
\end{matrix}\right.
\end{eqnarray}
where $\hat {A}^d(s)$, $\hat {B}^d(s)$, $\hat {C}^d(s)$, and $\hat {D}^d(s)$ are defined as
\begin{eqnarray}
\hat A^d(s)=\exp\biggl(\pi \lambda \left\{{\cal F}(s,R_\text{c}) -\frac{1}{2} {\cal F}(s,d+D)- \frac{1}{2}{\cal F}(s,\sqrt{D^2-d^2})\right\}\biggr),\hspace{100pt}
\end{eqnarray}
\begin{eqnarray}
\hat B^d(s)=\exp\left(\frac{\pi}{2} \lambda \Bigl\{{\cal F}(s,R_\text{c})- {\cal F}(s,d+D)\Bigr\}\right),\hspace{213pt}
\end{eqnarray}
\begin{eqnarray}
\hat C^d(s)=\exp\left(\lambda \Bigl\{ \varphi_0{\cal F}(s,R_\text{c}) + \left(\varphi_1(R_\text{c})-\varphi_0\right){\cal F}(s,R_1(\varphi_1(R_\text{c})))-\varphi_1(R_\text{c}) {\cal F}(s,d+D)\Bigr\}\right),
\end{eqnarray}
\begin{eqnarray}
\hat D^d(s)=\exp\left(\lambda \varphi_1(R_\text{c})\Bigl\{{\cal F}(s,R_\text{c})-{\cal F}(s,d+D)\Bigr\}\right).\hspace{190pt}
\end{eqnarray}

\end{corollary}
\begin{IEEEproof}
The proof follows the same approach as in Appendix A, except that the disk is replaced with the regions given in Fig. 4.
\end{IEEEproof}
\begin{corollary}
With uniform-selection, the LT of the interference is lower bounded by
\begin{eqnarray}
{\cal L}_{{\cal I}_\text{ub}}^d (s)= \left\{\begin{matrix}
\hat E^d(s) & \text{if} \hspace{+2pt}d \leq D, \\ 
\hat F^d(s) & \text{if} \hspace{+2pt}d > D.
\end{matrix}\right.
\end{eqnarray}
where $\hat E^d(s)$ and $\hat F^d(s)$ are defined as
\begin{eqnarray}
\hat E^d(s)= \frac{\exp\Bigl(-\frac{\lambda \pi}{2}\Bigl\{{\cal F}(s,\sqrt{D^2-d^2})+{\cal F}(s,D+d)\Bigr\}\Bigr)-\exp(-\lambda \pi D(D+d))}{(1-\exp(-\lambda \pi D(D+d)))\left(1-\frac{1}{2D(D+d)}\Bigl\{{\cal F}(s,\sqrt{D^2-d^2})+{\cal F}(s,D+d)\Bigr\}\right)},
\end{eqnarray}
\begin{eqnarray}
\hat F^d(s)= \frac{\exp\Bigl(- \lambda \varphi_0\Bigl\{{\cal F}(s,d+D)-{\cal F}(s,d-D)\Bigr\}\Bigr)-\exp(-4\lambda d D \varphi_0)}{(1-\exp(-4\lambda d D \varphi_0))\left(1-\frac{1}{4dD}\Bigl\{{\cal F}(s,d+D)-{\cal F}(s,d-D)\Bigr\}\right)}.\hspace{60pt}
\end{eqnarray}
\end{corollary}
\begin{IEEEproof}
The proof follows the same approach as in Appendix B, except that the disk is replaced with the regions given in Fig. 4.
\end{IEEEproof}

\section{Multi-Cluster Wireless Networks}
In this section, we extend our analysis to multi-cluster wireless networks. Closed-access and open-access receivers are investigated in Subsections IV.A and IV.B, respectively.
\subsection{Closed-Access Receivers}
For the analysis, we consider the reference closed-access receiver at the origin and add a cluster ${\Phi}_{\mathbf{x}_\text{o}}$ with intensity $\lambda$ over the disk $\mathbf{b}(\mathbf{x}_\text{o},D)$, where $\|\mathbf{x}_\text{o}\|$ has PDF given in (1), to the network. Thanks to Slivnyak's theorem [10, Thm. 8.10], this additional cluster and its receiver become the representative cluster and receiver under expectation over ${\Phi}$. This means this link's performance corresponds to the average performance of all links in any realization of the network. Therefore, the serving transmitter under the closest-selection strategy is
\begin{eqnarray}
{{\rm{\mathbf{x}}}}_\text{c} = \arg \mathop {\min }\limits_{{\mathbf{y}}\in \left\{{\Phi}_{\mathbf{x}_\text{o}}\mid n({\Phi}_{\mathbf{x}_\text{o}})>0\right\}} \| {{\mathbf{y}}} \|,
\end{eqnarray}
while with the uniform-selection strategy, the serving transmitter is found by
\begin{eqnarray}
{{\rm{\mathbf{x}}}}_\text{u} = \text{Unif}\left\{ {\Phi}_{\mathbf{x}_\text{o}}\mid n({\Phi}_{\mathbf{x}_\text{o}})>0\right\}.
\end{eqnarray}
Then, the SINR at the origin with distance ${\| \mathbf{x} \|}$ relative to the serving transmitter is
\begin{eqnarray}
\text{SINR} = \frac{{{h_\mathbf{x}}{{\| \mathbf{x} \|}^{ - \alpha }}}}{{{\sigma ^2} + {{\cal I}_{\text{intra}}} + {{\cal I}_{\text{inter}}}}},
\end{eqnarray}
where ${{\cal I}_{\text{intra}}} = \mathop \sum \limits_{\mathbf{y} \in {{\Phi} _{\mathbf{x}_\text{o}}}\backslash \left\{\mathbf{x}\right\}} {h_{\mathbf{y}}}{\| {\mathbf{y}} \|^{ - \alpha }}$ denotes the intra-cluster interference caused by the transmitters inside the representative cluster, and ${{\cal I}_{\text{inter}}} = \mathop  \sum \limits_{{{\mathbf{y}} \in {{\Phi} }}} {h_{{\mathbf{y}}}}{\| {{\mathbf{y}}} \|^{ - \alpha }}$ represents the inter-cluster interference caused by the transmitters outside the representative cluster. 

Given the distance between the receiver and the center of the representative cluster, i.e., $\|\mathbf{x}_{\text{o}}\|$, the distributions of the distances $R_\text{c}={\| \mathbf{x}_\text{c} \|}$ and $R_\text{u}={\| \mathbf{x}_\text{u} \|}$ are derived in Subsection III.B for the closest- and uniform-selection strategies, respectively. Also, the LT of the intra-cluster interference conditioned on $\|\mathbf{x}_\text{o}\|$, defined as the function ${\cal L}_{{\cal I}_{\text{intra}}}^{\|\mathbf{x}_\text{o}\|}$, is given by Theorems 1 and 2 for the closest-selection and uniform-selection strategies, relatively. The LT of the inter-cluster interference is also characterized in the following theorem.
\begin{theorem}
The LT of the inter-cluster interference ${\cal I}_{\text{inter}}$ is
\begin{eqnarray}
\hspace{-10pt}{\cal L}_{{\cal I}_\text{inter}} (s)=\exp\biggl(-2\pi \lambda_{\rm p} \biggr(\int\limits_{0}^{D}\Bigl\{1-\exp\left(-\lambda f(s,u)\right)\Bigr\}u\mathrm{d}u +\int\limits_{D}^{\infty}\Bigl\{1-\exp\left(-\lambda g(s,u)\right)\Bigr\}u\mathrm{d}u\biggr) \biggr),
\end{eqnarray}
where $f(s,u)$ and $g(s,u)$ are defined as
\begin{eqnarray}
f(s,u)=\int_{0}^{\pi}{\cal F}(s,{R_2(u,\theta) })\mathrm{d}\theta ,\hspace{115pt}
\end{eqnarray}
\begin{eqnarray}
g(s,u)=\int_{ 0}^{ \varphi_2(u)}\Bigl\{{\cal F}(s,{R_2(u,\theta) })-{\cal F}(s,{\hat R_2(u,\theta) })\Bigr\}\mathrm{d}\theta ,
\end{eqnarray}
where $\varphi_2 (u)= {\sin ^{ - 1}}\left( {\frac{D}{u}} \right)$, $R_2 (u,\theta)= u\cos \left( \theta  \right)+\sqrt {{D^2}-{u^2}{{\sin }^2}\left( \theta  \right) }$ and $\hat R_2 (u,\theta)= u\cos \left( \theta  \right)-\sqrt {D^2-{u^2}{{\sin }^2}\left( \theta  \right) }$.
\end{theorem}
\begin{IEEEproof}
See Appendix C.
\end{IEEEproof}

Using the expressions derived for the intra- and inter-cluster interferences, the coverage probability under the closest-selection strategy is calculated by deconditioning as
\begin{eqnarray}
{P}_{\rm C}^\text{ca-c} (\beta)= \mathbb{P}(n({\Phi}_{\mathbf{x}_\text{o}})>0)\mathbb{P}(\text{SINR}_\text{c} > \beta \mid n({\Phi}_{\mathbf{x}_\text{o}})>0)= (1-\exp(-\lambda \pi D^2))\times\nonumber\\\int_{0}^{\infty}\int_{0}^{\infty} {\mathbb{P}}\left( {\frac{{{h_\mathbf{x}}{{r}^{ - \alpha }}}}{{{\sigma ^2} + {{\cal I}_{\text{intra}}} + {{\cal I}_{\text{inter}}}}} > \beta } \right){f_{R_\text{c}}^{v}}\left( r \right) {f_{\|\mathbf{x}_\text{o}\|}}\left( v \right) \mathrm{d}r \mathrm{d} v, \hspace{+50pt}
\end{eqnarray}
where the conditional coverage probability is expressed as
\begin{eqnarray}
{\mathbb{P}}\left( {\frac{{{h_\mathbf{x}}{{r}^{ - \alpha }}}}{{{\sigma ^2} + {{\cal I}_{\text{intra}}} + {{\cal I}_{\text{inter}}}}} > \beta} \right) = \exp\left(-{{{\beta {\sigma ^2}}}} {r^\alpha }\right) {\cal L}_{{\cal I}_\text{c}}^{v}\left({{\beta {r^\alpha }}}|r\right){\cal L}_{{\cal I}_\text{inter}}\left({{\beta {r^\alpha }}}\right).
\end{eqnarray}
Then, according to the cases in Subsections III.B and C, (37) is given by
\begin{eqnarray}
{P}_{\rm C}^\text{ca-c}(\beta) =\int_{0}^{D} \int_{0}^{D-v} 2\pi \lambda r e^{-\lambda \pi r^2}\exp\left(-{{{\beta {\sigma ^2}}}} {r^\alpha }\right) {\cal L}_{\cal {I_{\text{c}}}}^v\left({{\beta {r^\alpha }}}|r\right){\cal L}_{\cal {I_{\text{inter}}}}\left({{\beta {r^\alpha }}}\right){f_{\|\mathbf{x}_\text{o}\|}}\left( v\right) \mathrm{d}r \mathrm{d}v\nonumber\hspace{+0pt}\\+ \int_{0}^{D}\int_{D-v}^{D+v} \lambda \frac{{\partial {\cal B}_v(r) }}{{\partial r}}{e^{ - \lambda {\cal B}_v(r)}}\exp\left(-{{{\beta {\sigma ^2}}}} {r^\alpha }\right) {\cal L}_{\cal I_{\text{c}}}^v\left({{\beta {r^\alpha }}}|r\right){\cal L}_{\cal {I_{\text{inter}}}}\left({{\beta {r^\alpha }}}\right){f_{\|\mathbf{x}_\text{o}\|}}\left( v \right) \mathrm{d}r\mathrm{d} v\nonumber\hspace{+4pt}\\+\int_{D}^{\infty} \int_{v -D}^{D+v} \lambda \frac{{\partial {\cal B}_v(r) }}{{\partial r}}{e^{ - \lambda {\cal B}_v(r)}}\exp\left(-{{{\beta {\sigma ^2}}}} {r^\alpha }\right) {\cal L}_{\cal I_{\text{c}}}^v\left({{\beta {r^\alpha }}}|r\right){\cal L}_{\cal {I_{\text{inter}}}}\left({{\beta {r^\alpha }}}\right) {f_{\|\mathbf{x}_\text{o}\|}}\left( v \right) \mathrm{d}r\mathrm{d}v.\hspace{0pt}
\end{eqnarray}

Finally, the coverage probability with the uniform-selection strategy can be derived with the same procedure as in (37)-(39). It is given by
\begin{eqnarray}
{P}_\text{\rm C}^\text{ca-u}(\beta)  =(1-e^{-\lambda \pi D^2})\biggl\{\int_{0}^{D} \int_{0}^{D-v} \frac{{2r}}{D^2}\exp\left(-{{{\beta {\sigma ^2}}}} {r^\alpha }\right) {\cal L}_{\cal {I_{\text{u}}}}^v\left({{\beta {r^\alpha }}}\right){\cal L}_{\cal {I_{\text{inter}}}}\left({{\beta {r^\alpha }}}\right){f_{\|\mathbf{x}_\text{o}\|}}\left( v\right) \mathrm{d}r \mathrm{d}v\nonumber\hspace{+0pt}\\+ \int_{0}^{D}\int_{D-v}^{D+v} \frac{{1}}{\pi D^2} \frac{{\partial {\cal B}_v(r) }}{{\partial r}}\exp\left(-{{{\beta {\sigma ^2}}}} {r^\alpha }\right) {\cal L}_{\cal I_{\text{u}}}^v\left({{\beta {r^\alpha }}}\right){\cal L}_{\cal {I_{\text{inter}}}}\left({{\beta {r^\alpha }}}\right){f_{\|\mathbf{x}_\text{o}\|}}\left( v \right) \mathrm{d}r\mathrm{d} v\nonumber\hspace{+17pt}\\+\int_{D}^{\infty} \int_{v -D}^{D+v} \frac{{1}}{\pi D^2} \frac{{\partial {\cal B}_v(r) }}{{\partial r}}\exp\left(-{{{\beta {\sigma ^2}}}} {r^\alpha }\right) {\cal L}_{\cal I_{\text{u}}}^v\left({{\beta {r^\alpha }}}\right){\cal L}_{\cal {I_{\text{inter}}}}\left({{\beta {r^\alpha }}}\right) {f_{\|\mathbf{x}_\text{o}\|}}\left( v \right) \mathrm{d}r\mathrm{d}v\biggr\}.\hspace{3.5pt}
\end{eqnarray}
\subsection{Open-Access Receivers}
Here, we only consider the closest-selection strategy. Note that there is no uniform-selection strategy in this case, since the number of transmitters is infinite. Therefore, the serving transmitter is
\begin{eqnarray}
{{\rm{\mathbf{x}}}}_\text{t} = \arg \mathop {\min }\limits_{{\mathbf{y}} \in {\Phi} } \| {{\mathbf{y}}} \|.
\end{eqnarray}
The distribution of the distance from the reference open-access receiver at the origin, i.e., $R_\text{t}={\| \mathbf{x}_\text{t} \|}$,  is given in the following theorem.
\begin{theorem}
The cumulative distribution function (CDF) of ${R_\text{t}}$ is
\begin{eqnarray}
F_{R_\text{t}}(r)= \left\{\begin{matrix}
1-\exp\biggl(-2\pi \lambda_{\rm p}\biggl\{\left(1-\exp \left( - \lambda \pi r^2 \right)\right) \frac{(D-r)^2}{2}\hspace{+0pt}\\+\int_{D-r}^{D+r}\left(1-\exp \left( - \lambda {\cal B}_u(r) \right)\right)u\mathrm{d}u\biggr\}\biggr)\hspace{30pt}  & \hspace{+20pt}\text{if} \hspace{+5pt}0\leq r<D, \\ 
1-\exp\biggl(-2\pi \lambda_{\rm p}\biggl\{\left(1-\exp \left( - \lambda \pi D^2 \right)\right) \frac{(r-D)^2}{2}\hspace{+0pt}\\+\int_{r-D}^{D+r}\left(1-\exp \left( - \lambda {\cal B}_u(r) \right)\right)u\mathrm{d}u\biggr\}\biggr)\hspace{30pt}  & \text{if} \hspace{+5pt}r\geq D.
\end{matrix}\right.
\end{eqnarray}
\end{theorem}
\begin{IEEEproof}
See Appendix D.
\end{IEEEproof}
Note that the contact distribution function of MCP is also derived in [28] using the probability generating functional (PGFL) of PCPs in [10, Cor. 4.13]. However, our approach is more tractable and leads to the result (42), which is much easier to numerically evaluate than [28, Thm. 1].  

By taking the derivative of $F_{R_\text{t}}(r)$, using Leibniz integral rule, and some simplifications, the PDF of $R_\text{t}$ can be obtained as 
\begin{eqnarray}
{f_{R_\text{t}}}\left( r \right) = \left\{\begin{matrix}
2\pi \lambda_{\rm p}\biggl\{\pi \lambda r e^{-\lambda \pi r^2}(D-r)^2+\int_{D-r}^{D+r}\lambda \frac{{\partial {\cal B}_u(r)}}{{\partial r}} e^{-\lambda {\cal B}_u(r)}u\mathrm{d}u\biggr\}\hspace{0pt}\\\times\exp\biggl(-2\pi \lambda_{\rm p}\biggl\{\left(1-\exp \left( - \lambda \pi r^2 \right)\right) \frac{(D-r)^2}{2}\hspace{47pt}\\+\int_{D-r}^{D+r}\left(1-\exp \left( - \lambda {\cal B}_u(r) \right)\right)u\mathrm{d}u\biggr\}\biggr)\hspace{71pt}  & \hspace{+20pt}\text{if} \hspace{+5pt}0\leq r<D, \\ 
2\pi \lambda_{\rm p}\biggl\{\int_{r-D}^{D+r}\lambda \frac{{\partial {\cal B}_u(r)}}{{\partial r}} e^{-\lambda {\cal B}_u(r)}u\mathrm{d}u\biggr\}\hspace{105pt}\\\times\exp\biggl(-2\pi \lambda_{\rm p}\biggl\{ \left(1-\exp \left( - \lambda \pi D^2 \right)\right) \frac{(r-D)^2}{2}\hspace{42pt}\\+\int_{r-D}^{D+r}\left(1-\exp \left( - \lambda {\cal B}_u(r) \right)\right)u\mathrm{d}u\biggr\}\biggr)\hspace{70pt}  & \text{if} \hspace{+5pt}r\geq D.
\end{matrix}\right.
\end{eqnarray}
The SINR at the reference open-access receiver located at the origin is
\begin{eqnarray}
\text{SINR}_\text{t} = \frac{{{h_{\mathbf{x}_\text{t}}}{{R_\text{t}}^{ - \alpha }}}}{{{\sigma ^2} + {{\cal I}_{\text{t}}}}},
\end{eqnarray}
where ${{\cal I}_{\text{t}}} = \mathop \sum \limits_{\mathbf{y} \in {{\Phi}}\backslash \left\{\mathbf{x}_\text{t}\right\}} {h_{\mathbf{y}}}{\| {\mathbf{y}} \|^{ - \alpha }}$ denotes the total interference caused by all transmitters except for the serving transmitter. The LT of the total interference ${\cal I}_{\text{t}}$ conditioned on the serving distance $R_\text{t}$ is characterized in the following theorem. 
\begin{theorem}
Conditioned on $R_\text{t}$, the LT of the total interference ${\cal I}_{\text{t}}$ is
\begin{eqnarray}
{\cal L}_{{\cal I}_\text{t}} (s|R_\text{t}) = \left\{\begin{matrix}
\hspace{-100pt}\exp\biggl(-2\pi \lambda_{\rm p}\biggl\{ \int_{ 0}^{D-R_\text{t}}\left\{1-{A}^u(s)\right\}u\mathrm{d}u\hspace{-3pt}\\+\int_{ D-R_\text{t}}^{\sqrt{D^2+R_\text{t}^2}}\left\{1-{B}^u(s)\right\}u\mathrm{d}u+\int_{ \sqrt{D^2+R_\text{t}^2}}^{D+R_\text{t}}\left\{1-{C}^u(s)\right\}u\mathrm{d}u\hspace{-23pt}\\+\int_{D+R_\text{t}}^{\infty}\left\{1-\exp\left(-\lambda g(s,u)\right)\right\}u\mathrm{d}u\biggr\}\biggr)\hspace{+74pt}  & \hspace{0pt}\text{if} \hspace{+5pt}0\leq R_\text{t}<D, \\ 
\exp\biggl(-2\pi \lambda_{\rm p}\biggl\{\int_{ R_\text{t}-D}^{\sqrt{D^2+R_\text{t}^2}}\left\{1-{B}^u(s)\right\}u\mathrm{d}u\hspace{80pt}\\+\int_{ \sqrt{D^2+R_\text{t}^2}}^{D+R_\text{t}}\left\{1-{C}^u(s)\right\}u\mathrm{d}u\hspace{+125pt}\\+\int_{D+R_\text{t}}^{\infty}\left\{1-\exp\left(-\lambda g(s,u)\right)\right\}u\mathrm{d}u\biggr\}\biggr) \hspace{74pt} &\hspace{-17pt} \text{if} \hspace{+5pt}R_\text{t}\geq D,
\end{matrix}\right.
\end{eqnarray}
where ${A}^u(s)$, ${B}^u(s)$ and ${C}^u(s)$ are defined for a $u$ in (12), (13) and (14), respectively. Also, $g(s,u)$ is given in (36).
\end{theorem}
\begin{IEEEproof}
See Appendix E.
\end{IEEEproof}
Using Theorem 5 and (43) and the same procedure as in Subsection III.C, the coverage probability of the reference open-access receiver in multi-cluster wireless networks is found as
\begin{eqnarray}
{P}_{\rm C}^\text{oa} (\beta)= \int_{0}^{\infty} \exp\left(-{{{\beta {\sigma ^2}}}} {r^\alpha }\right) {\cal L}_{{\cal I}_\text{t}}\left({{\beta {r^\alpha }}}|r\right) f_{R_\text{t}}(r) \mathrm{d}r.
\end{eqnarray}
\section{Numerical Results and Discussion}
In this section, we provide analytical results for specific scenarios of single- and multi-cluster wireless networks. In addition, we discuss the results and provide key design insights.
\subsection{Single-Cluster Wireless Networks}
We consider a scenario of single-cluster wireless networks in which the transmitters are distributed according to an FHPPP with intensity $\lambda = 0.01 \hspace{+2pt}\text{m}^{-2}$ in a disk with radius $D = 15 \hspace{+2pt}\text{m}$ and evaluate the coverage probability results derived in Subsection III.C. Also, we evaluate the spectral efficiency, defined as $\tau = \mathbb{E}\left\{\text{log}(1+\text{SINR})\right\}$. The variance of the additive white Gaussian noise is set to $\sigma^2 = 0.0001$. We further define the normalized (relative) distance $\delta = \frac{d}{D}$. In the following, we study the impact of the path loss exponent and the distance of the receiver from the center of the disk on the coverage probability and the spectral efficiency. We also investigate the tightness of the bounds derived in Subsection III.D.

\textit{Effect of path loss exponent:} The coverage probability as a function of the minimum required SINR $\beta$ is plotted in Fig. 5 for the closest- and uniform-selection strategies, considering $\delta = \frac{2}{3}$ and $\frac{4}{3}$ and $\alpha = 3$ and $4$. It is observed that the coverage probability is improved by increasing the path loss exponent. However, with the uniform-selection, a higher path loss exponent has a degrading effect when $\beta$ is lower than $-5$ dB. That is because the SINR exhibits a tradeoff as $\alpha$ increases. The power of both the desired and the interfering signals decrease, which can lead to an increase or decrease in the SINR, depending on $\beta$ and the other parameters. At a target coverage probability of $0.8$, the (horizontal) gap between uniform- and closest-selection is about $20$ dB for $\alpha=3$ and $25$ dB for $\alpha=4$ in the case $\delta = \frac{2}{3}$ and $12$ dB for $\alpha=3$ and $18$ dB for $\alpha=4$ in the case $\delta = \frac{4}{3}$.

\textit{Effect of receiver distance from the center:} The coverage probability as a function of the normalized distance $\delta$ is studied in Fig. 6 for both selection strategies, $\alpha = 4$ and $\beta = -5$ and $0$ dB. It is observed that, depending on $\beta$, there is an optimal value for the distance of the receiver, about $2D$ for uniform-selection at $\beta = -5$ dB and $0.8D$ for other cases, in terms of the coverage probability. This is due to the fact that the SINR has a tradeoff since the power of both the desired and the interfering signals decrease as the distance of the receiver to the center of the disk increases.

\textit{Tightness of the bounds:} The tightness of the bounds on coverage probability derived in Subsection III.D is evaluated in Fig. 7 in the cases with different selection strategies and $\alpha = 4$ and $\delta = \frac{2}{3}$ and $\frac{4}{3}$. As observed, the bounds tightly approximate the performance in a broad range of SINR thresholds $\beta$ and for different positions of the receiver inside or outside the disk.

\textit{Spectral efficiency:} The spectral efficiency as a function of $\delta$ is shown in Fig. 8 for both selection strategies and $\alpha = 3$ and $4$. We analytically obtain the spectral efficiency from the coverage probability as, e.g., [8, Thm. 3]\footnote{Note that this is an upper bound on the actual spectral efficiency, since this formulation assumes that the transmitter knows all the fading coefficients between all other transmitters and the receiver, which would be a very generous assumption. A tighter lower bound could be found using the approach described in [33].}
\begin{eqnarray}
\tau^q = \int_{0}^{\infty} \mathbb{P}(\text{log}\left(1+\text{SINR}_q)>t\right)\mathrm{d}t = \int_{0}^{\infty}\frac{1}{\text{ln}2}\frac{{P}_{\rm C}^q\left(t\right)}{t+1} \mathrm{d}t \hspace{+5pt}\text{bits/channel use},\ q = \left\{\text{c},\text{u}\right\}.
\end{eqnarray}

As observed, closest-selection achieves a much higher spectral efficiency than uniform-selection for different receiver distances. Also, there is a crossing point, whereby the spectral efficiency improves as $\alpha$ increases before reaching a distance for the receiver location outside the disk. Then, at higher distances, the reverse happens. This is intuitive because, at lower distances, the power of the interfering signals, which decreases with a higher path loss exponent, is the dominant factor on the SINR. On the other hand, at higher distances, the power of the desired signal, which increases with a smaller path loss exponent, dominates. Also, there may be an optimal value for the distance of the receiver in terms of the spectral efficiency. This is the result of the coverage probability behavior with the distance. 
\begin{figure}[tb!]
\centering
%coverage_pathloss_2.eps
\includegraphics[width =3.2in]{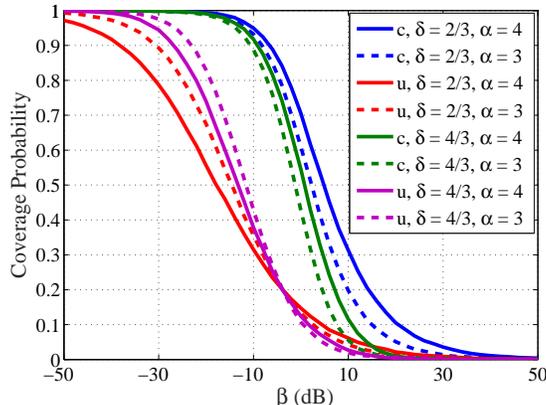}
\vspace{-15pt}
\caption{Coverage probability as a function of the SINR threshold $\beta$ with $\delta = \frac{2}{3}$ and $\frac{4}{3}$. c and u denote closest- and uniform-selection, respectively.}
\vspace{-15pt}
\end{figure}

\begin{figure}[t!]
%distance_2.eps
\centering
\center
\vspace{-1ex}
\includegraphics[width =3.2in]{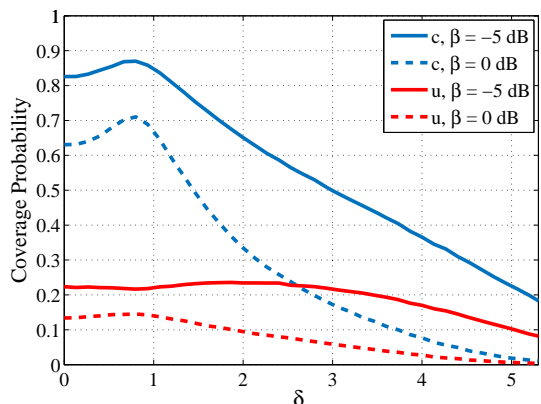}
\vspace{-15pt}
\caption{Coverage probability as a function of normalized distance $\delta$ with $\alpha = 4$.}
\vspace{-15pt}
\end{figure}
\begin{figure}[t!]
%bound3.eps
\centering
\center
\vspace{-1ex}
\includegraphics[width =3.2in]{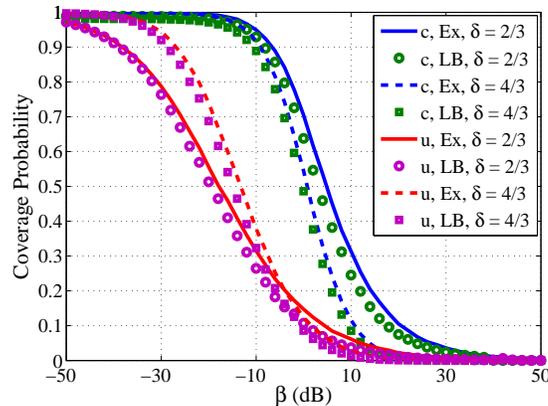}
\vspace{-15pt}
\caption{On the tightness of the coverage probability lower bounds, $\alpha = 4$. Ex and LB denote the exact result and the lower bound, respectively.}
\vspace{-15pt}
\end{figure}
\begin{figure}[t!]
%rate_2.eps
\centering
\center
\vspace{-1ex}
\includegraphics[width =3.2in]{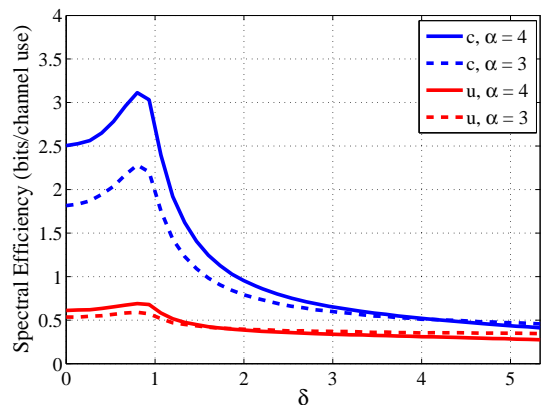}
\vspace{-15pt}
\caption{Spectral efficiency as a function of normalized distance $\delta$.}
\vspace{-15pt}
\end{figure}

\begin{figure}[t!]
%coverage_rate2.eps
\centering
\center
\vspace{-1ex}
\includegraphics[width =3.2in]{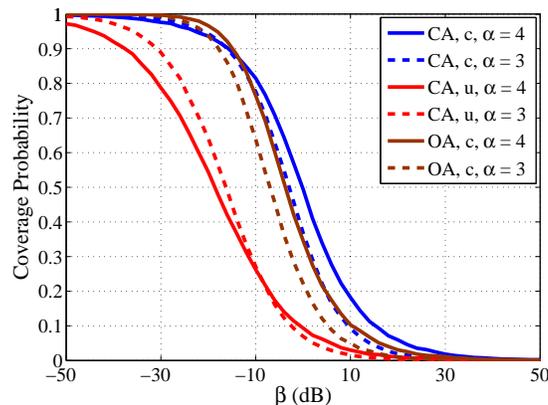}
\vspace{-15pt}
\caption{Coverage probability of the closed-access (CA) and open-access (OA) receiver as a function of threshold $\beta$ with $\delta_{\rm c} = \frac{2}{3}$.}
\vspace{-15pt}
\end{figure}

\begin{figure}[t!]
%var_2.eps
\centering
\center
\vspace{-1ex}
\includegraphics[width =3.2in]{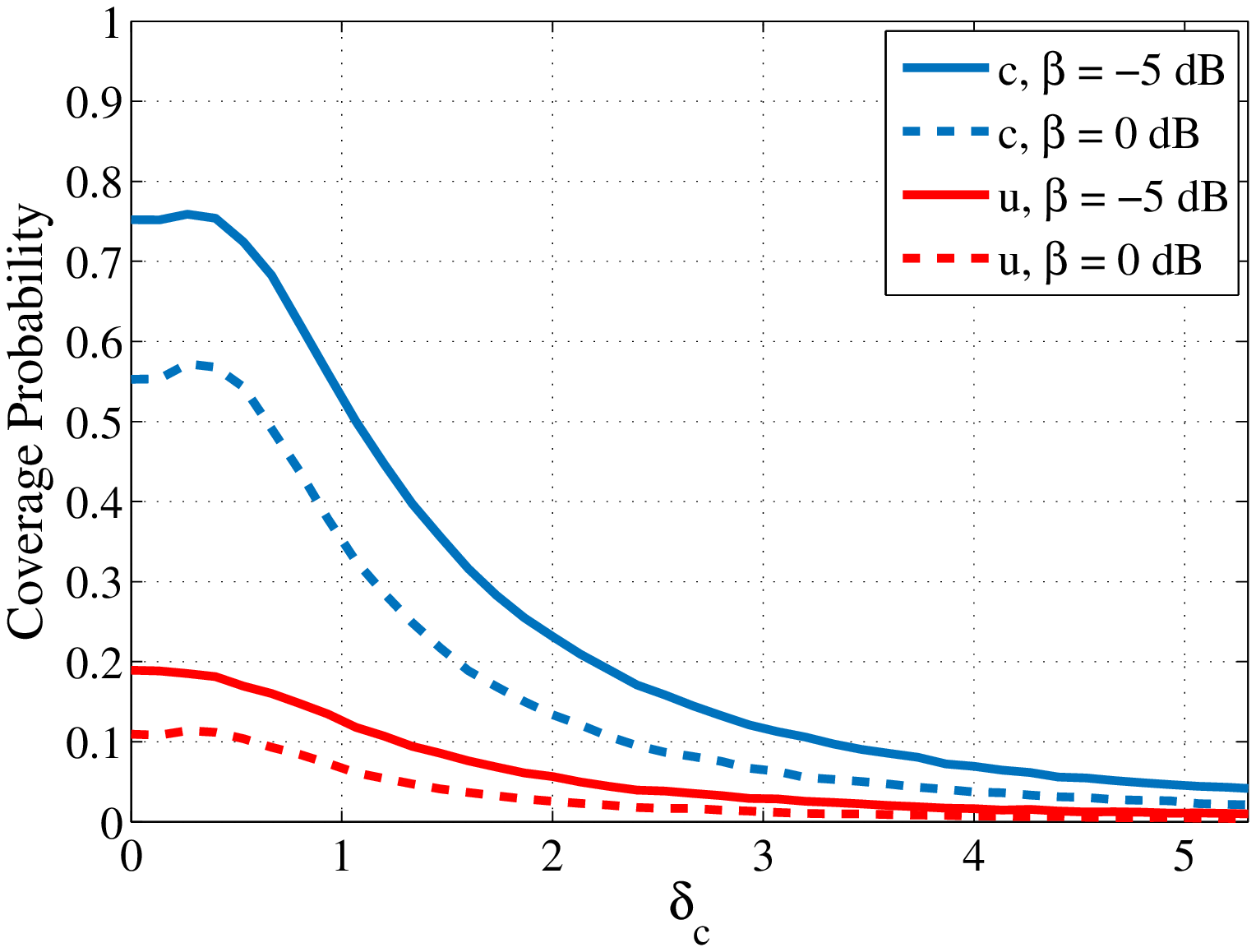}
\vspace{-15pt}
\caption{Coverage probability of the closed-access receiver as a function of normalized standard deviation $\delta_{\rm c}$ with $\alpha = 4$.}
\vspace{-20pt}
\end{figure}

\subsection{Multi-Cluster Wireless Networks}
Here, we consider a multi-cluster wireless network where transmitters form an MCP with intensity $\lambda = 0.01 \hspace{+2pt}\text{m}^{-2}$ inside disks, all with equal radius $D = 15 \hspace{+2pt}\text{m}$, whose centers follow an HPPP with intensity $\lambda_{\rm p} = 0.0004 \hspace{+2pt}\text{m}^{-2}$. Closed-access receivers are distributed with normal distribution with variance $\sigma_\text{c}^2$ around the cluster centers of transmitters. We further define normalized standard deviation $\delta_\text{c} = \frac{\sigma_\text{c}}{D}$. In the following, the effects of the path loss exponent and variance of the normal distribution of closed-access receivers are assessed on the theoretical results derived in Subsections IV.A and IV.B.

\textit{Effect of path loss exponent:} In Fig. 9, the coverage probability as a function of $\beta$ is plotted for the closest- and uniform-selection strategies in the case of closed-access receivers with $\delta_\text{c} = \frac{2}{3}$. In this figure, the case of open-access receivers with the closest-selection strategy is also considered. Here, the results are presented for $\alpha = 3$ and $4$. As observed, a higher path loss exponent improves the coverage probability in practical minimum required SINRs.

\textit{Effect of variance of the normal distribution:} The coverage probability as a function of the normalized standard deviation $\delta_\text{c}$ is shown in Fig. 10. Here, the results are presented for the closed-access receivers, considering $\alpha = 4$ and $\beta = -5$ and $0$ dB. It is observed that the coverage probability decreases as the standard deviation (or variance) of the normal distribution increases. It is intuitive because the probability of the event that the receiver is located farther from the representative transmitters increases with $\sigma_\text{c}$.
\section{Conclusion}
In this paper, we developed a comprehensive tractable framework for the modeling and analysis of single- and multi-cluster wireless networks. Suitable for different wireless applications, we considered two strategies for a reference receiver to select a serving transmitter over a single-cluster wireless network: closest-selection and uniform-selection. Considering two types of receivers---closed-access receivers and open-access receivers---we then extended our modeling to multi-cluster wireless networks that are composed of distributed single-cluster wireless networks.

Using tools from stochastic geometry, we derived exact expressions for the coverage probability in the cases with different transmitter selection strategies and types of receivers in single- and multi-cluster wireless networks. We also proposed tight closed-form expressions bounding the coverage probability in the case of single-cluster wireless networks. Our analysis revealed that a higher path loss exponent improves the performance except when the receiver is outside and relatively far from its cluster of transmitters. In addition, although an increase in the distance of the receiver to the center of the cluster typically degrades the performance, there exists a location with optimal performance for the receiver. Also, the closest-selection strategy significantly outperforms the uniform-selection strategy.
%Also, the distributions of the distances and the LT of the interferences were derived. 

\appendices

\section{Proof of Theorem 1}
The LT of the interference, given the serving distance $R_\text{c}$, is calculated as
\begin{eqnarray}
{\cal L}_{{\cal I}_\text{c}}^d (s|R_\text{c})=\mathbb{E}\left\{\exp\left(-s\mathop \sum \nolimits_{\mathbf{y} \in {\Phi} \backslash \left\{\mathbf{x}_\text{c}\right\}} {h_\mathbf{y}}{{\| \mathbf{y} \|}^{-\alpha}}\right)\mid n({\Phi})>0\right\}\nonumber\hspace{130pt}\\=\mathbb{E}\Biggl\{\mathop \prod \limits_{\mathbf{y} \in {\Phi} \backslash \left\{\mathbf{x}_\text{c}\right\}} \exp\left(-s {h_\mathbf{y}}{{\| \mathbf{y} \|}^{-\alpha}}\right)\mid n({\Phi})>0\Biggr\}\stackrel{(a)}{=}\mathbb{E}\Biggl\{\mathop \prod \limits_{\mathbf{y} \in {\Phi} \backslash \left\{\mathbf{x}_\text{c}\right\}}\frac{1}{{1 + s{{\| \mathbf{y} \|}^{ - \alpha }}}}\mid n({\Phi})>0\Biggr\}\nonumber\\\stackrel{(b)}{=}\exp\left(-\lambda \int_{{\cal A}\backslash \mathbf{b}(\mathbf{o},R_\text{c})}^{} \left(1- \frac{1}{{1 + s{{\| \mathbf{y} \|}^{ - \alpha }}}}\right)\mathrm{d}\mathbf{y}\right),\hspace{190.5pt}
\end{eqnarray}
where $(a)$ is obtained by $h_\mathbf{y} \sim \exp(1)$ and $(b)$ follows from the PGFL of the PPP [10, Thm. 4.9] and the fact that interfering nodes are farther away than $R_\text{c}$. Having a non-empty intersection between ${\cal A}$ and $\mathbf{b}(\mathbf{o},R_\text{c})$, to compute (48), there are two types for $d<D$ (Case 1) and two types for $d>D$ (Case 2) to convert from Cartesian to polar coordinates. In fact, each type denotes a special form of ${\cal A}\backslash \mathbf{b}(\mathbf{o},R_\text{c})$ that can be represented by polar coordinates uniquely.

\textbf{Case 1:} We have two different types for the case $d<D$ as follows.

Type 1: If ${\cal A}\cap \mathbf{b}(\mathbf{o},R_\text{c}) = \mathbf{b}(\mathbf{o},R_\text{c})$ as given in Fig. 2(b), i.e., $0\leq R_\text{c}<D-d$, then
\begin{eqnarray}
{\cal L}_{{\cal I}_\text{c}}^d (s|R_\text{c}) = \exp\biggl(-\lambda \int\limits_{0}^{2\pi}\int\limits_{0}^{R_1(\theta)} \left(1- \frac{1}{{1 + s{{x}^{ - \alpha }}}}\right)x\mathrm{d}x\mathrm{d}\theta +\lambda \int\limits_{ 0}^{2\pi}\int\limits_{0}^{R_\text{c}} \left(1- \frac{1}{{1 + s{{x}^{ - \alpha }}}}\right)x\mathrm{d}x\mathrm{d}\theta \biggr).
\end{eqnarray}
This can be simplified to
\begin{eqnarray}
{\cal L}_{{\cal I}_\text{c}}^d (s|R_\text{c}) =\exp\left(2\pi \lambda \int_{0}^{R_\text{c}}\frac{x}{{1 + \frac{{{x^\alpha }}}{s}}}\mathrm{d}x-\lambda \int_{0}^{2\pi}\int_{0}^{R_1(\theta)} \frac{x}{{1 + \frac{{{x^\alpha }}}{s}}}\mathrm{d}x\mathrm{d}\theta \right)\nonumber\\ \stackrel{(c)}{=}\exp\left(\pi \lambda {\cal F}(s,R_\text{c})-{\lambda} \int_{ 0}^{\pi}{\cal F}(s,R_1(\theta))\mathrm{d}\theta \right),\hspace{+66pt}
\end{eqnarray}
%where ${\cal Z}(\alpha,s,R)=\frac{1}{\alpha }{s^{\frac{2}{\alpha }}}{\cal B}\left( {\frac{2}{\alpha },1 - \frac{2}{\alpha }} \right) - \frac{{s{R^{2 - \alpha }}}}{{\alpha  - 2}}{}_2^{}{F_1}\left( {1,1 - \frac{2}{\alpha };2 - \frac{2}{\alpha }; - \frac{s}{{{R^\alpha }}}} \right)$. Equality $(c)$ follows from replacing $r^{\alpha}$ with $u$ and using the beta function ${\cal B}[.]$ and Gauss hypergeometric function ${}_2^{}{F_1}[.]$. 
where $(c)$ follows from replacing $x^{\alpha}$ with $u$ and calculating
the corresponding integral based on the formula [32, (3.194.1)] which uses the Gauss hypergeometric function ${}_2{F_1}$. 

Type 2: If ${\cal A}\cap \mathbf{b}(\mathbf{o},R_\text{c}) \ne \mathbf{b}(\mathbf{o},R_\text{c}) $ as given in Fig. 2(a), i.e., $D-d \leq R_\text{c}<D+d$, then
\begin{eqnarray}
{\cal L}_{{\cal I}_\text{c}}^d (s|R_\text{c}) = \exp\left(-\lambda \int_{ -\varphi_1(R_\text{c})}^{\varphi_1(R_\text{c})}\int_{R_\text{c}}^{{R}_1(\theta)} \left(1- \frac{1}{{1 + s{{x}^{ - \alpha }}}}\right)x\mathrm{d}x\mathrm{d}\theta \right) \hspace{0pt}\nonumber\\=\exp\biggl(-\lambda \int_{0}^{\varphi_1(R_\text{c})}\Bigl\{{\cal F}(s,R_1(\theta)) - {\cal F}(s,R_\text{c}) \Bigr\}\mathrm{d}\theta \biggr). \hspace{+17pt}
\end{eqnarray}

\textbf{Case 2:} As observed from Fig. 2(c) for the case $d>D$, we have two different types of the LT of the interference depending on whether or not, at an angle, the lower boundary of $\cal A$ within the two tangent lines crossing the origin is included in the boundary of ${\cal A}\backslash \mathbf{b}(\mathbf{o},R_\text{c})$.

Type 1: Since the intersection angle $\varphi_1(r)$ at $r = \sqrt{d^2-D^2}$ is equal to the angle of the tangent lines $\varphi_0$, if $d-D<R_\text{c}<\sqrt{d^2-D^2}$, we have 
\begin{eqnarray}
{\cal L}_{{\cal I}_\text{c}}^d (s|R_\text{c}) = \exp\biggl(-\lambda \biggl\{\int_{-\varphi_0 }^{-\varphi_1(R_\text{c})}\int_{\hat R_1(\theta)}^{R_1(\theta)} \left(1- \frac{1}{{1 + s{{x}^{ - \alpha }}}}\right)x\mathrm{d}x\mathrm{d}\theta\nonumber\hspace{+130pt}\\+\int_{ -\varphi_1(R_\text{c})}^{ \varphi_1(R_\text{c})}\int_{R_\text{c}}^{R_1(\theta)} \left(1- \frac{1}{{1 + s{{x}^{ - \alpha }}}}\right)x\mathrm{d}x\mathrm{d}\theta+\int_{ \varphi_1(R_\text{c})}^{\varphi_0}\int_{\hat R_1(\theta)}^{R_1(\theta)} \left(1- \frac{1}{{1 + s{{x}^{ - \alpha }}}}\right)x\mathrm{d}x\mathrm{d}\theta\biggr\} \biggr)\nonumber\hspace{+5pt}\\ =\exp\biggl(-\lambda \int\limits_{\varphi_1(R_\text{c})}^{\varphi_0}\Bigl\{{\cal F}(s,R_1(\theta))-{\cal F}(s,\hat R_1(\theta))\Bigr\}\mathrm{d}\theta-{\lambda} \int\limits_{0}^{\varphi_1(R_\text{c})}\Bigl\{{\cal F}(s,R_1(\theta))-{\cal F}(s,R_\text{c})\Bigr\}\mathrm{d}\theta \biggr).
\end{eqnarray}

Type 2: If $\sqrt{d^2-D^2}<R_\text{c}<d+D$, then 
\begin{eqnarray}
{\cal L}_{{\cal I}_\text{c}}^d (s|R_\text{c}) = \exp\left(-\lambda \int_{ -\varphi_1(R_\text{c})}^{ \varphi_1(R_\text{c})}\int_{R_\text{c}}^{R_1(\theta)} \left(1- \frac{1}{{1 + s{{x}^{ - \alpha }}}}\right)x\mathrm{d}x\mathrm{d}\theta \right) \nonumber\hspace{+0pt}\\ =\exp\biggl(-{\lambda} \int_{0}^{\varphi_1(R_\text{c})}\Bigl\{{\cal F}(s,R_1(\theta))-{\cal F}(s,R_\text{c})\Bigr\}\mathrm{d}\theta \biggr).\hspace{+17pt}
\end{eqnarray}
\section{Proof of Theorem 2}
The LT of the interference is
\begin{eqnarray}
{\cal L}_{{\cal I}_\text{u}}^d (s)=\mathbb{E}\left\{\exp\left(-s\mathop \sum \nolimits_{\mathbf{y} \in {\Phi} \backslash \left\{\mathbf{x}_\text{u}\right\}} {h_\mathbf{y}}{{\| \mathbf{y} \|}^{-\alpha}}\right)\mid n({\Phi})>0\right\}\nonumber\hspace{150pt}\\=\mathbb{E}\Biggl\{\mathop \prod \limits_{\mathbf{y} \in {\Phi} \backslash \left\{\mathbf{x}_\text{u}\right\}} \exp\left(-s {h_\mathbf{y}}{{\| \mathbf{y} \|}^{-\alpha}}\right)\mid n({\Phi})>0\Biggr\}\stackrel{(a)}{=}\mathbb{E}\Biggl\{\mathop \prod \limits_{\mathbf{y} \in {\Phi} \backslash \left\{\mathbf{x}_\text{u}\right\}}\frac{1}{{1 + s{{\| \mathbf{y} \|}^{ - \alpha }}}}\mid n({\Phi})>0\Biggr\}\nonumber\\
\stackrel{(b)}{=}\mathbb{E}\Biggl\{\left(\int_{0}^{\infty}\frac{1}{{1 + s{{x}^{ - \alpha }}}}f_{R_\text{u}}(x)\mathrm{d}x\right)^{n({\Phi})-1}\mid n({\Phi})>0\Biggr\}\nonumber\hspace{170pt}\\\stackrel{(c)}{=}\mathop \sum \limits_{k=1}^{\infty}\frac{\exp(-\lambda \pi D^2)(\lambda \pi D^2)^k}{k! (1-\exp(-\lambda \pi D^2))}\left(\int_{0}^{\infty}\frac{1}{{1 + s{{x}^{ - \alpha }}}}f_{R_\text{u}}(x)\mathrm{d}x\right)^{k-1}\nonumber\hspace{133pt}
\\=\frac{1}{(1-\exp(-\lambda \pi D^2))\int_{0}^{\infty}\frac{1}{{1 + s{{x}^{ - \alpha }}}}f_{R_\text{u}}(x)\mathrm{d}x}\times\nonumber\hspace{205pt}\\\left(\mathop \sum \limits_{k=0}^{\infty}\frac{\exp(-\lambda \pi D^2)(\lambda \pi D^2)^k}{k! }\left(\int_{0}^{\infty}\frac{1}{{1 + s{{x}^{ - \alpha }}}}f_{R_\text{u}}(x)\mathrm{d}x\right)^{k}-\exp(-\lambda \pi D^2)\right)\nonumber\hspace{50pt}
\end{eqnarray}
\begin{eqnarray}
\stackrel{(d)}{=}\frac{\exp(-\lambda \pi D^2)}{(1-\exp(-\lambda \pi D^2))\int\limits_{0}^{\infty}\frac{1}{{1 + s{{x}^{ - \alpha }}}}f_{R_\text{u}}(x)\mathrm{d}x}\Biggl(\exp\Biggl(\lambda \pi D^2\int\limits_{0}^{\infty}\frac{1}{{1 + s{{x}^{ - \alpha }}}}f_{R_\text{u}}(x)\mathrm{d}x\Biggr)-1\Biggr),\hspace{+0pt}
\end{eqnarray}
where ${f_{R_\text{u}}^d}$ is the PDF of $R_\text{u}$ obtained from the distributions in (9)-(10) and $(a)$ is found by $h_\mathbf{y} \sim \exp(1)$. Also, $(b)$ follows from the fact that conditioned on $n(\Phi)$, $\Phi$ is a BPP where the distance $\|\mathbf{y}\|$ of each point is i.i.d. with distribution $f_{R_\text{u}}^d$, $(c)$ follows from $\mathbb{P}(n({\Phi})=k\mid n({\Phi})>0)=\frac{\exp(-\lambda \pi D^2)(\lambda \pi D^2)^k}{k! (1-\exp(-\lambda \pi D^2))}$, and $(d)$ is found by the moment-generating function (MGF) of a Poisson random variable with mean $\lambda \pi D^2$ and some simplifications. According to (9)-(10), there are two cases to compute the integral:

\textbf{Case 1}: If $d \leq D$, then
\begin{eqnarray}
\int_{0}^{\infty}\frac{1}{{1 + s{{x}^{ - \alpha }}}}f_{R_\text{u}}(x)\mathrm{d}x = \int_{0}^{D-d}\frac{1}{{1 + s{{x}^{ - \alpha }}}}\frac{2x}{D^2}\mathrm{d}x+\int_{D-d}^{D+d}\frac{1}{{1 + s{{x}^{ - \alpha }}}}\frac{{1}}{\pi D^2} \frac{{\partial {\cal B}_d(x) }}{{\partial x}}\mathrm{d}x.
\end{eqnarray}

\textbf{Case 2}: If $d > D$, then
\begin{eqnarray}
\int_{0}^{\infty}\frac{1}{{1 + s{{x}^{ - \alpha }}}}f_{R_\text{u}}(x)\mathrm{d}x =\int_{d-D}^{D+d}\frac{1}{{1 + s{{x}^{ - \alpha }}}}\frac{{1}}{\pi D^2} \frac{{\partial {\cal B}_d(x) }}{{\partial x}}\mathrm{d}x.
\end{eqnarray}

\section{Proof of Theorem 3}
The LT of the inter-cluster interference is
\begin{eqnarray}
\hspace{-10pt}{\cal L}_{{\cal I}_\text{inter}} (s)=\mathbb{E}\Biggl\{\exp\Biggl(-s \mathop \sum \limits_{\mathbf{x}\in {\Phi}_{\rm p}}\mathop \sum \limits_{{{\mathbf{y}} \in {{\Phi} _\mathbf{x}}}} {h_{{\mathbf{y}}}}{\| {{\mathbf{y}}} \|^{ - \alpha }}\Biggr)\Biggr\}=\mathbb{E}\Biggl\{\mathop \prod \limits_{{\mathbf{x}\in {\Phi}_{\rm p}}}\mathop \prod \limits_{\mathbf{y} \in {\Phi}_\mathbf{x} } \exp\left(-s {h_{{\mathbf{y}}}}{\| {{\mathbf{y}}} \|^{ - \alpha }}\right)\Biggr\}\nonumber\hspace{13pt}\\\hspace{0pt}\stackrel{(a)}{=}\mathbb{E}\Biggl\{\mathop \prod \limits_{{\mathbf{x}\in {\Phi}_{\rm p}}}\mathop \prod \limits_{\mathbf{y} \in {\Phi}_\mathbf{x} }\frac{1}{{1 + s{{\| \mathbf{y} \|}^{ - \alpha }}}}\Biggr\}\stackrel{(b)}{=}\mathbb{E}\left\{\mathop \prod \limits_{\mathbf{x} \in {\Phi}_{\rm p}}\mathop \prod \limits_{\mathbf{z} \in {\Phi}_\text{o} }\frac{1}{{1 + s{{\| \mathbf{z}+\mathbf{x} \|}^{ - \alpha }}}}\right\} \hspace{+55pt}\nonumber\\\stackrel{(c)}{=}\mathbb{E}\left\{\mathop \prod \limits_{\mathbf{x} \in {\Phi}_{\rm p} }\exp\left(-\lambda \int_{\mathbf{b}(\mathbf{o},D)}^{} \left(1- \frac{1}{{1 + s{{\| \mathbf{z} +\mathbf{x} \|}^{ - \alpha }}}}\right)\mathrm{d}\mathbf{z}\right)\right\}\hspace{+86.5pt}\nonumber
\end{eqnarray}
\begin{eqnarray}
\stackrel{(d)}{=}\exp\left(-\lambda_{\rm p} \int_{\mathbb{R}^2}{}\left\{1-\exp\left(-\lambda \int_{\mathbf{b}(\mathbf{o},D)}^{} \left(1- \frac{1}{{1 + s{{\| \mathbf{z} +\mathbf{x} \|}^{ - \alpha }}}}\right)\mathrm{d}\mathbf{z}\right)\right\}\mathrm{d}\mathbf{x}\right),\hspace{-31pt}
\end{eqnarray}
where $(a)$ follows from $h_\mathbf{y} \sim \exp(1)$, and $(b)$ comes from the fact that $\mathbf{y} \in {\Phi}_\mathbf{x}$ can be written as $\mathbf{y}= \mathbf{z}+\mathbf{x}$, where $\mathbf{z}$ is taken from the FHPPP ${\Phi}_\text{o}$ with intensity $\lambda$ over $\mathbf{b}(\mathbf{o},D)$ and $\mathbf{x} \in {\Phi}_{\rm p}$. Also, both $(c)$ and $(d)$ follow from the PGFL of the PPP. In order to convert the inner integral from Cartesian to polar coordinates,  there are two cases:

\textbf{Case 1}: If $\| \mathbf{x} \| \leq D$, then
\begin{eqnarray}
f(s,\| \mathbf{x} \|)  =  \int_{\mathbf{b}(\mathbf{o},D)}^{} \left(1- \frac{1}{{1 + s{{\| \mathbf{z} +\mathbf{x} \|}^{ - \alpha }}}}\right)\mathrm{d}\mathbf{z}=\int_{0}^{ 2\pi}\int_{0}^{R_2 (\| \mathbf{x} \|,\theta)} \left(1- \frac{1}{{1 + s{{x}^{ - \alpha }}}}\right)x\mathrm{d}x\mathrm{d}\theta \nonumber\\= \int_{0}^{\pi}{\cal F}(s,R_2 (\| \mathbf{x} \|,\theta))\mathrm{d}\theta,\hspace{78pt}
\end{eqnarray}
where $R_2 (\| \mathbf{x} \|,\theta)= \sqrt {{D^2}-{\| \mathbf{x} \|^2}{{\sin }^2}\left( \theta  \right) }+\| \mathbf{x} \|\cos \left( \theta  \right)$.

\textbf{Case 2}: If $\| \mathbf{x} \| > D$, then
\begin{eqnarray}
g(s, \| \mathbf{x} \|)  = \int_{\mathbf{b}(\mathbf{o},D)}^{} \left(1- \frac{1}{{1 + s{{\| \mathbf{z} +\mathbf{x} \|}^{ - \alpha }}}}\right)\mathrm{d}\mathbf{z}= \int\limits_{ -\varphi_2 (\| \mathbf{x} \|)}^{ \varphi_2 (\| \mathbf{x} \|)}\int\limits_{\hat R_2 (\| \mathbf{x} \|,\theta)}^{R_2 (\| \mathbf{x} \|,\theta)} \left(1- \frac{1}{{1 + s{{x}^{ - \alpha }}}}\right)x\mathrm{d}x\mathrm{d}\theta\nonumber\\= \int_{ 0}^{ \varphi_2(\| \mathbf{x} \|)}\Bigl\{{\cal F}(s,R_2 (\| \mathbf{x} \|,\theta))-{\cal F}(s,\hat R_2 (\| \mathbf{x} \|,\theta))\Bigr\}\mathrm{d}\theta ,\hspace{128pt}
\end{eqnarray}
where $\varphi_2 (\| \mathbf{x} \|)= {\sin ^{ - 1}}\left( {\frac{D}{\| \mathbf{x} \|}} \right)$ and $\hat R_2 (\| \mathbf{x} \|,\theta)= \| \mathbf{x} \|\cos \left( \theta  \right)-\sqrt {{D^2}-{\| \mathbf{x} \|^2}{{\sin }^2}\left( \theta  \right) }$.
Therefore, converting the outer integral from Cartesian to polar coordinates, the final result can be written as
\begin{eqnarray}
\hspace{-4pt}{\cal L}_{{\cal I}_\text{inter}} (s)=\exp\biggl(-2\pi \lambda_\text{c} \biggl(\int\limits_{0}^{D}\Bigl\{1-\exp\left(-\lambda f(s,u)\right)\Bigr\}u\mathrm{d}u+\int\limits_{D}^{\infty}\Bigl\{1-\exp\left(-\lambda g(s,u)\right)\Bigr\}u\mathrm{d}u\biggr) \biggr).
\end{eqnarray}

\section{Proof of Theorem 4}
Let us define $R_{\mathbf{x}}=\mathop{\min }\limits_{{\mathbf{y}} \in {\Phi}_{{\mathbf{x}}} } \| {{\mathbf{y}}} \|$, which is the distance from the reference receiver to its closest transmitter in the cluster with parent point $\mathbf{x}\in {\Phi}_{\rm p}$. Then, the CDF of $R_\text{t}=\mathop{\min }\limits_{{\mathbf{x}}\in {\Phi}_{\rm p}} R_{{\mathbf{x}}}$ is
\begin{eqnarray}
F_{R_\text{t}}(r)=\mathbb{P}\left( {R_\text{t}< r} \right) =1 - \mathbb{P}\left( {\mathop {\min }\limits_{{\mathbf{x}}\in {\Phi}_{\rm p}}R_{{\mathbf{x}}}> r} \right) \stackrel{(a)}{=} 1-\mathop {\lim }\limits_{\rho\to \infty }\mathbb{E}\left\{\mathbb{P}\Bigl( {\mathop {\min }\limits_{{\mathbf{x}}\in {\Psi}_{\rho}}R_{{\mathbf{x}}}> r}\mid n({\Psi}_{\rho}) \Bigr)\right\}\nonumber\hspace{+4pt}\\\stackrel{(b)}{=} 1 - \mathop {\lim }\limits_{\rho \to \infty }\mathbb{E}\Biggl\{\mathbb{E}\Biggl\{\mathop \prod \limits_{{\mathbf{x}}\in {\Psi}_{\rho}} \mathbb{P}\left( {R_\mathbf{x} > r}\right) \mid n({\Psi}_{\rho})\Biggr\} \Biggr\}\nonumber\hspace{175pt}\\ \stackrel{(c)}{=} 1 - \mathop {\lim }\limits_{\rho \to \infty }\mathbb{E}\Biggl\{\mathbb{E}\Biggl\{\mathop \prod \limits_{{\mathbf{x}}\in {\Psi}_{\rho}} \bigl(\mathbb{P}\left(n({\Phi}_\mathbf{x})=0\right)\mathbb{P}\left( {R_\mathbf{x} > r} \mid n({\Phi}_\mathbf{x})=0\right)\nonumber\hspace{95pt}\\
+ \mathbb{P}\left(n({\Phi}_\mathbf{x})>0\right)\mathbb{P}\left( {R_\mathbf{x} > r} \mid n({\Phi}_\mathbf{x})>0\right)\bigr) \mid n({\Psi}_{\rho})\Biggr\}\Biggr\}\nonumber\hspace{130pt}
\end{eqnarray}
\begin{eqnarray}
\stackrel{(d)}{=} 1 - \mathop {\lim }\limits_{\rho \to \infty }\mathbb{E}\Biggl\{\mathbb{E}\Biggl\{ \mathop \prod \limits_{\mathbf{x} \in {\Psi}_{\rho}}\exp(-\lambda \pi D^2)+(1-\exp(-\lambda \pi D^2)){\cal E} (\mathbf{x})\mid n({\Psi}_{\rho})\Biggr\}\Biggr\} \nonumber\hspace{+72pt}\\\stackrel{(e)}{=}1-\mathop {\lim }\limits_{\rho \to \infty }\mathbb{E}\Biggl\{\left(\frac{1}{\pi {\rho}^2}\int_{\mathbf{b}(\mathbf{o},\rho)}^{}\exp(-\lambda \pi D^2)+(1-\exp(-\lambda \pi D^2)){\cal E} (\mathbf{x})\mathrm{d}\mathbf{x}\right)^{n({\Psi}_{\rho})}\Biggr\}\nonumber\hspace{+53pt}\\\stackrel{(f)}{=}1-\mathop {\lim }\limits_{\rho \to \infty }\exp\Biggl(-\lambda_{\rm p}\pi {\rho}^2\Biggl(1-\frac{1}{\pi {\rho}^2}\int\limits_{\mathbf{b}(\mathbf{o},\rho)}^{}\exp\left(-\lambda \pi D^2\right)+\left(1-\exp\left(-\lambda \pi D^2\right)\right){\cal E} (\mathbf{x})\mathrm{d}\mathbf{x}\Biggr)\Biggr)\nonumber\hspace{+0pt}\\=1-\exp\left(-\lambda_{\rm p}\int_{\mathbb{R}^2}^{}\left(1-\exp\left(-\lambda \pi D^2\right)\right)(1-{\cal E} (\mathbf{x}))\mathrm{d}\mathbf{x}\right),\hspace{+157pt}
\end{eqnarray}
where ${\Psi}_{\rho} = {\Phi}_{\rm p}\cap \mathbf{b}(\mathbf{o},\rho)$ and $(a)$ follows from $\mathop {\lim }\limits_{\rho \to \infty } {\Psi}_{\rho}= {\Phi}_{\rm p}$. Then, $(b)$ is due to the facts that $R_\mathbf{x}$ is a function of $\mathbf{x}\in {\Psi}_{\rho}$ and ${\Psi}_{\rho}$ conditioned on $n({\Psi}_{\rho})$ is a BPP. Also, $(c)$ is obtained by conditioning on the existence of a transmitter inside each cluster, and $(d)$ comes from Subsection III.B, whereby ${\cal E} (\mathbf{x})$ is defined as
\begin{eqnarray}
{\cal E}(\mathbf{x}) = \frac{\exp\left( { - {\rm{\lambda }}\left| {\rm{\cal C}}_\mathbf{x} \right|} \right)-\exp({ - \lambda \pi D^2})}{1-\exp(-\lambda \pi D^2)}
\end{eqnarray}
where $|{\cal C}_\mathbf{x}|$ is the area of the intersection between $\mathbf{b}(\mathbf{o},r)$ and $\mathbf{b}(\mathbf{x},D)$. Also, note that $\\\mathbb{P}\left( {{R_\mathbf{x}} > r} \mid n({\Phi}_\mathbf{x})=0\right) = 1, \hspace{+4pt}\text{for}\hspace{+5pt} 0<r<\infty$. In addition, $(e)$ follows from the fact that the points $\mathbf{x}$ are i.i.d. uniformly
distributed on $\mathbf{b}(\mathbf{o},\rho)$, and $(f)$ is found by the MGF of the number of points of ${\Psi}_{\rho}$, which is Poisson with mean ${\lambda}_{\rm p}\pi{\rho}^2$. 

According to Fig. 3, there are two cases for $|{\cal C}_\mathbf{x}|$:

\textbf{Case 1:} If $\| {{\mathbf{x}}} \| \le D$, then
\begin{eqnarray}
\left| \cal C_\mathbf{x} \right| = \left\{ {\begin{array}{*{20}{c}}
{\pi {r^2} \hspace{+100pt} \| {{\mathbf{x}}} \|<D-r,}\\
{{\cal B_{\|\mathbf{x}\|}}(r) \hspace{+50pt} \| {{\mathbf{x}}} \|>r-D \hspace{+5pt}\text{and}\hspace{+5pt} \| {{\mathbf{x}}} \| \ge D-r,\hspace{-50pt}}\\
{\pi D^2 \hspace{+100pt} \| {{\mathbf{x}}} \| \leq r-D,}
\end{array}} \right. 
\end{eqnarray}

\textbf{Case 2:} If $\| {{\mathbf{x}}} \| > D$, then
\begin{eqnarray}
\left| \cal C_\mathbf{x} \right| = \left\{ {\begin{array}{*{20}{c}}
{0 \hspace{+110pt} \| {{\mathbf{x}}} \|>r+D,}\\
{{\cal B_{\|\mathbf{x}\|}}(r) \hspace{+50pt} \| {{\mathbf{x}}} \|>r-D \hspace{+5pt}\text{and}\hspace{+5pt}\| {{\mathbf{x}}} \|\leq r+D,\hspace{-50pt}}\\
{\pi D^2 \hspace{+100pt} \| {{\mathbf{x}}} \|\leq r-D,}
\end{array}} \right. 
\end{eqnarray}
where $\mathcal{B}_{\|\mathbf{x}\|}(r)$ is given in (5). Therefore, converting (61) from Cartesian to polar coordinates according to (62)-(64), the CDF of $R_\text{t}$ is found as
\begin{eqnarray}
F_{R_\text{t}}(r)= 1-\exp\biggl\{-2\pi \lambda_{\rm p}\int_{0}^{D-r}\left(1-\exp(-\lambda \pi r^2)\right)u\mathrm{d}u+\int_{D-r}^{D}\left(1-\exp \left( - \lambda {\cal B}_u(r) \right)\right)u\mathrm{d}u\nonumber\\+\int_{D}^{D+r}\left(1-\exp \left( - \lambda {\cal B}_u(r) \right)\right)u\mathrm{d}u+\int_{D+r}^{\infty}\left(1-\exp \left(  0 \right)\right)u\mathrm{d}u\biggr\}\biggr),\hspace{81pt}
\end{eqnarray}
if $0\leq r<D,$
\begin{eqnarray}
F_{R_\text{t}}(r)= 1-\exp\biggl(-2\pi \lambda_{\rm p}\biggl\{\int_{0}^{r-D}(1-\exp(-\lambda \pi D^2))\left(1-0\right)u\mathrm{d}u\nonumber\hspace{+57pt}\\+\int_{r-D}^{D}\left(1-\exp \left( - \lambda {\cal B}_u(r) \right)\right)u\mathrm{d}u+\int_{D}^{D+r}\left(1-\exp \left( - \lambda {\cal B}_u(r) \right)\right)u\mathrm{d}u\nonumber\\+\int_{D+r}^{\infty}\left(1-\exp \left(  0 \right)\right)u\mathrm{d}u\biggr\}\biggr),\hspace{189pt}
\end{eqnarray}
if $D\leq r< 2D,$ and 
\begin{eqnarray}
F_{R_\text{t}}(r)= 1-\exp\biggl(-2\pi \lambda_{\rm p}\biggl\{\int_{0}^{D}(1-\exp(-\lambda \pi D^2))\left(1-0\right)u\mathrm{d}u\nonumber\hspace{+102pt}\\+\int_{D}^{r-D}(1-\exp(-\lambda \pi D^2))\left(1-0\right)u\mathrm{d}u+\int_{r-D}^{D+r}\left(1-\exp \left( - \lambda {\cal B}_u(r) \right)\right)u\mathrm{d}u\nonumber\\+\int_{D+r}^{\infty}\left(1-\exp \left(  0 \right)\right)u\mathrm{d}u\biggr\}\biggr),\hspace{222pt}
\end{eqnarray}
if $r\geq 2D$. Here, though derived differently, (66) and (67) leads to the same expression. The final result can be obtained by some simplifications.

\section{Proof of Theorem 5}
The LT of the total interference, conditioned on the serving distance $R_\text{t}$, is
\begin{eqnarray}
{\cal L}_{{\cal I}_\text{t}} (s|R_\text{t})\stackrel{(a)}{=}\mathbb{E}\Biggl\{\exp\Biggl(-s \mathop \sum \limits_{{{\mathbf{y}} \in {{\Phi} } \backslash \mathbf{b}(\mathbf{o},R_\text{t})}} {h_{{\mathbf{y}}}}{\| {{\mathbf{y}}} \|^{ - \alpha }}\Biggr)\Biggr\}\hspace{+208pt}\nonumber\\=\mathbb{E}\Biggl\{\mathop \prod \limits_{{\mathbf{x}\in {\Phi}_{\rm p}}}\mathop \prod \limits_{\mathbf{y} \in {\Phi}_\mathbf{x} \backslash \mathbf{b}(\mathbf{o},R_\text{t})} \exp\left(-s {h_{{\mathbf{y}}}}{\| {{\mathbf{y}}} \|^{ - \alpha }}\right)\Biggr\}\stackrel{(b)}{=}\mathbb{E}\Biggl\{\mathop \prod \limits_{{\mathbf{x}\in {\Phi}_{\rm p}}}\mathop \prod \limits_{\mathbf{y} \in {\Phi}_\mathbf{x} \backslash \mathbf{b}(\mathbf{o},R_\text{t})}\frac{1}{{1 + s{{\| \mathbf{y} \|}^{ - \alpha }}}}\Biggr\}\hspace{+4pt}\nonumber\\ \stackrel{(c)}{=}\mathbb{E}\left\{\mathop \prod \limits_{\mathbf{x} \in {\Phi}_{\rm p} }\exp\left(-\lambda \int_{\mathbf{b}(\mathbf{x},D)\backslash \mathbf{b}(\mathbf{o},R_\text{t})}^{} \left(1- \frac{1}{{1 + s{\| \mathbf{y} \|}^{ - \alpha }}}\right)\mathrm{d}\mathbf{y}\right)\right\}\hspace{+90pt}\nonumber\\\stackrel{(d)}{=}\exp\left(-\lambda_{\rm p} \int_{\mathbb{R}^2}{}\left\{1-\exp\left(-\lambda \int_{\mathbf{b}(\mathbf{x},D)\backslash \mathbf{b}(\mathbf{o},R_\text{t})}^{} \left(1- \frac{1}{{1 + s{{\| \mathbf{y}\|}^{ - \alpha }}}}\right)\mathrm{d}\mathbf{y}\right)\right\}\mathrm{d}\mathbf{x}\right),\hspace{5pt}
\end{eqnarray}
where $(a)$ follows from the fact that all transmitters with distance more than $R_\text{t}$ are interferers, $(b)$ comes from $h_\mathbf{y} \sim \exp(1)$, and both of $(c)$ and $(d)$ follow from the PGFL. According to Theorems 1 and 3, there are the following cases for the inner integral.

\textbf{Case 1:} If $\| {{\mathbf{x}}} \| \le D$ and $\| {{\mathbf{x}}} \|<D-R_\text{t}$, then
\begin{eqnarray}
 \exp\left(-\lambda  \int_{\mathbf{b}(\mathbf{x},D)\backslash \mathbf{b}(\mathbf{o},R_\text{t})}^{} \left(1- \frac{1}{{1 + s{{\| \mathbf{y} \|}^{ - \alpha }}}}\right)\mathrm{d}\mathbf{y}\right) = {A}^{\| {{\mathbf{x}}} \|}(s).
\end{eqnarray}

\textbf{Case 2:} If $\| {{\mathbf{x}}} \| \le D$ and $\| {{\mathbf{x}}} \|>R_\text{t}-D$ and $\| {{\mathbf{x}}} \|\ge D-R_\text{t}$, then
\begin{eqnarray}
\exp\left(-\lambda  \int_{\mathbf{b}(\mathbf{x},D)\backslash \mathbf{b}(\mathbf{o},R_\text{t})}^{} \left(1- \frac{1}{{1 + s{{\| \mathbf{y} \|}^{ - \alpha }}}}\right)\mathrm{d}\mathbf{y}\right) ={B}^{\| {{\mathbf{x}}} \|}(s).
\end{eqnarray}

\textbf{Case 3:} If $\| {{\mathbf{x}}} \| > D$ and $\| {{\mathbf{x}}} \|>\sqrt{R_\text{t}^2+D^2}$ and $\| {{\mathbf{x}}} \| \le R_\text{t}+D$, then
\begin{eqnarray}
 \exp\left(-\lambda \int_{\mathbf{b}(\mathbf{x},D)\backslash \mathbf{b}(\mathbf{o},R_\text{t})}^{} \left(1- \frac{1}{{1 + s{{\left| {\left| \mathbf{y} \right|} \right|}^{ - \alpha }}}}\right)\mathrm{d}\mathbf{y}\right) = {C}^{\| {{\mathbf{x}}} \|}(s).
\end{eqnarray}

\textbf{Case 4:} If $\| {{\mathbf{x}}} \| > D$ and $\| {{\mathbf{x}}} \|>R_\text{t}-D$ and $\| {{\mathbf{x}}} \| \le \sqrt{R_\text{t}^2+D^2}$, then
\begin{eqnarray}
 \exp\left(-\lambda \int_{\mathbf{b}(\mathbf{x},D)\backslash \mathbf{b}(\mathbf{o},R_\text{t})}^{} \left(1- \frac{1}{{1 + s{{\| \mathbf{y} \|}^{ - \alpha }}}}\right)\mathrm{d}\mathbf{y}\right) = {B}^{\| {{\mathbf{x}}} \|}(s).
\end{eqnarray}

\textbf{Case 5:} If $\| {{\mathbf{x}}} \| > R_\text{t}+D$, then
\begin{eqnarray}
\exp\left(-\lambda  \int_{\mathbf{b}(\mathbf{x},D)\backslash \mathbf{b}(\mathbf{o},R_\text{t})}^{} \left(1- \frac{1}{{1 + s{{\| \mathbf{y} \|}^{ - \alpha }}}}\right)\mathrm{d}\mathbf{y}\right) = \exp\left(-\lambda g(s,\| {{\mathbf{x}}} \|)\right) .
\end{eqnarray}

\textbf{Case 6:} If $\| {{\mathbf{x}}} \| < R_\text{t}-D$, then
\begin{eqnarray}
\exp\left(-\lambda  \int_{\mathbf{b}(\mathbf{x},D)\backslash \mathbf{b}(\mathbf{o},R_\text{t})}^{} \left(1- \frac{1}{{1 + s{{\| \mathbf{y}\|}^{ - \alpha }}}}\right)\mathrm{d}\mathbf{y}\right) = 0.
\end{eqnarray}
Therefore, converting the outer integral from Cartesian to polar coordinates according to (69)-(74), the LT can be given by
\begin{eqnarray}
 {\cal L}_{{\cal I}_\text{t}} (s|R_\text{t}) = \exp\biggl(-2\pi \lambda_{\rm p}\biggl( \int_{ 0}^{D-R_\text{t}}\left\{1-{A}^u(s)\right\}u\mathrm{d}u\hspace{+149pt}\nonumber\\+\int_{ D-R_\text{t}}^{D}\left\{1-{B}^u(s)\right\}u\mathrm{d}u+\int_{ D}^{\sqrt{D^2+R_\text{t}^2}}\left\{1-{B}^u(s)\right\}u\mathrm{d}u\hspace{+63pt}\nonumber\\+\int_{ \sqrt{D^2+R_\text{t}^2}}^{D+R_\text{t}}\left\{1-{C}^u(s)\right\}u\mathrm{d}u+\int_{D+R_\text{t}}^{\infty}\left\{1-\exp\left(-\lambda g(s,u)\right)\right\}u\mathrm{d}u\biggr)\biggr), \hspace{-1pt}
 \end{eqnarray}
if $0 \le R_\text{t}< D,$
 \begin{eqnarray}
 {\cal L}_{{\cal I}_\text{t}} (s|R_\text{t}) = \exp\biggl(-2\pi \lambda_{\rm p}\biggl(\int_{ 0}^{R_\text{t}-D}\left\{1-\exp(0)\right\} u\mathrm{d}u\hspace{+169pt}\nonumber\\+\int_{R_\text{t}-D}^{D}\left\{1-{B}^u(s)\right\}u\mathrm{d}u+\int_{ D}^{\sqrt{D^2+R_\text{t}^2}}\left\{1-{B}^u(s)\right\}u\mathrm{d}u\hspace{+87pt}\nonumber\\+\int_{ \sqrt{D^2+R_\text{t}^2}}^{D+R_\text{t}}\left\{1-{C}^u(s)\right\}u\mathrm{d}u+\int_{D+R_\text{t}}^{\infty}\left\{1-\exp\left(-\lambda g(s,u)\right)\right\}u\mathrm{d}u\biggr)\biggr),\hspace{+24pt}
 \end{eqnarray}
if $D\leq R_\text{t}< 2D,$ and
 \begin{eqnarray}
 {\cal L}_{{\cal I}_\text{t}} (s|R_\text{t}) = \exp\biggl(-2\pi\lambda_{\rm p}\biggl(\int_{ 0}^{R_\text{t}-D}\left\{1-\exp(0)\right\}u\mathrm{d}u+\int_{ R_\text{t}-D}^{\sqrt{D^2+R_\text{t}^2}}\left\{1-{B}^u(s)\right\}u\mathrm{d}u\nonumber\hspace{+0pt}\\+\int_{ \sqrt{D^2+R_\text{t}^2}}^{D+R_\text{t}}\left\{1-{C}^u(s)\right\}u\mathrm{d}u+\int_{D+R_\text{t}}^{\infty}\left\{1-\exp\left(-\lambda g(s,u)\right)\right\}u\mathrm{d}u\biggr)\biggr),\hspace{+5pt}
 \end{eqnarray}
if $R_\text{t}\geq 2D$. The final result can be obtained by some simplifications.

\end{document}